\documentstyle[preprint,tighten,aps,prc,epsfig]{revtex} 
\def\sst#1{{\scriptscriptstyle #1}}

\def\notv{{\not\! v}}

\def\sstw{{\sin^2\theta_\sst{W}}}

\def\sst#1{{\scriptscriptstyle #1}}

\def\GMS{{G_\sst{M}^{(s)}}}

\def\lamchi{{\Lambda_\chi}} 
 
\def\qw0{{Q_\sst{W}^0}} 
\def\qwp{{Q_\sst{W}^P}} 
\def\qwn{{Q_\sst{W}^N}}

\def\alr{{A_\sst{LR}}}

\def\mn{{m_\sst{N}}}

\def\mz{{M_\sst{Z}}}

\def\ratez{{R_\sst{A}^{T=0}}} 
\def\rateo{{R_\sst{A}^{T=1}}} 
 
\def\mk{{m_\sst{K}}}

\def\mpis{{m_\pi^2}}

\def\GMn{{G_\sst{M}^n}} 
\def\GMp{{G_\sst{M}^p}} 
\def\GAp{{G_\sst{A}^p}}

\def\RAp{{R_\sst{A}^p}}

\def\lamchi{{\Lambda_\chi}} 
\def\lamchis{{\Lambda_\chi^2}} 
 
\begin{document} 
 
\begin{titlepage} 
 
\begin{center} 
 
{\large{\bf The Nucleon Anapole Moment and Parity-Violating $ep$ Scattering}} 
 
\vspace{1.2cm} 
 
Shi-Lin Zhu$^a$, S.J. Puglia$^a$, B.R. Holstein$^c$, M. J. 
Ramsey-Musolf$^{a,b}$ 
 
\vspace{0.8cm} 
 
$^a$ Department of Physics, University of Connecticut, 
Storrs, CT 06269 USA\\ 
$^b$ Theory Group, Thomas Jefferson National Laboratory, Newport News, 
VA 23606 USA\\ 
$^c$ Department of Physics and Astronomy, Universityof Massachusetts, 
Amherst, MA 01003 USA 
 
\end{center} 
 
\vspace{1.0cm} 
 
\begin{abstract} 
Parity-violating (PV) interactions among quarks in the nucleon induce a PV 
$\gamma NN$ coupling, or 
anapole moment (AM). We compute electroweak gauge-independent contributions 
to the AM through 
${\cal O}(1/\lamchis)$ in chiral perturbation theory. We estimate 
short-distance PV effects using 
resonance saturation. The AM contributions to PV electron-proton scattering 
slightly enhance the 
axial vector radiative corrections, $R_A^p$, over the scale implied by the 
Standard Model when weak 
quark-quark interactions are neglected. We estimate the theoretical 
uncertainty associated with the 
AM contributions to $R_A^p$ to be large, and discuss the implications for 
the interpretation PV of 
$ep$ scattering. 
 
\vskip 0.5 true cm 
PACS Indices: 21.30.+y, 13.40.Ks, 13.88.+e, 11.30.Er 
 
\end{abstract} 
 
\vspace{2cm} 
\vfill 
\end{titlepage} 
 
\pagenumbering{arabic} 
\section{Introduction} 
\label{sec1} 
 
The SAMPLE collaboration at MIT-Bates has recently reported a value for the 
strange-quark magnetic form factor measured using backward angle 
parity-violating 
(PV) electron-proton scattering \cite{Spa00}: 
\begin{equation} 
\label{eq:gmsep} 
\GMS(Q^2=0.1\ {\hbox{GeV}}^2/c^2) = 0.61\pm 0.27 \pm 0.19\ \ \ , 
\end{equation} 
where the first error is experimental and the second is theoretical. The 
dominant 
contribution to the theoretical error is uncertainty associated with radiative 
corrections to the axial vector term in the backward angle left-right 
asymmetry $\alr$\cite{MRM94}: 
\begin{equation} 
\label{eq:epasym} 
\alr\propto \qwp+\qwn{\GMn\over\GMp} +Q_\sst{W}^{(0)}{\GMS\over\GMp} - 
(1-4\sstw)\sqrt{1+1/\tau}{\GAp\over\GMp} \ \ \ , 
\end{equation} 
where $\qwp$ and $\qwn$ are the proton and neutron weak charges, respectively, 
$Q_\sst{W}^{(0)}$ is the SU(3)-singlet weak charge\footnote{Note that in 
Ref. \cite{MRM94}, the 
weak charges are denoted $\xi_\sst{V}^{p,n, (0)}$ .}, $\theta_\sst{W}$ is the 
weak mixing angle, and $\tau=Q^2/4 M_n^2$. The axial form factor is 
normalized at 
the photon point as 
\begin{equation} 
\GAp (0) = -g_\sst{A} [1+\RAp] \ \ \ 
\end{equation} 
where $g_A=1.267 \pm 0.004$ \cite{pdg} is the nucleon's axial charge as 
measured in 
neutron $\beta$-decay and $\RAp$ denotes process-dependent electroweak 
radiative 
corrections to the $V(e)\times A(p)$ scattering amplitude. 
 
The radiative correction $\RAp$ is the subject of the present study. It was 
first analyzed in Ref. \cite{mike} and found to be large, negative in sign, and 
plagued by considerable theoretical uncertainty. Generally, $\RAp$ contains two 
classes of contributions. The first involve electroweak radiative 
corrections to 
the elementary $V(e)\times A(q)$ amplitudes, where $q$ is any one of the 
quarks in the proton. These radiative corrections, referred to henceforth 
as \lq\lq 
one-quark" radiative corrections, are calculable in the Standard Model. 
They contain 
little theoretical uncertainty apart from the gentle variation with Higgs 
mass and 
long-distance QCD effects involving light-quark loops in the $Z-\gamma$ mixing 
tensor. The one-quark contributions can be large, due to the absence from 
loops of the small 
$(1-4\sstw)$ factor appearing at tree level (see Eq. (\ref{eq:epasym}) ) 
and the 
presence of large logarithms of the type $\ln (m_q/\mz)$. 
 
A second class of radiative corrections, which we refer to as \lq\lq 
many-quark" corrections, involve weak interactions among quarks in the 
proton. In this 
paper, we focus on those many-quark corrections which generate an axial vector 
coupling of the photon to the proton (see Figure 1). This axial vector 
$pp\gamma$ 
interaction, also known as the anapole moment (AM), has the form 
\begin{equation} 
\label{eq:anapole} 
{\cal L}^\sst{AM}={e\over \Lambda_{\chi}^2}\bar N (a_s+a_v\tau_3 ) 
\gamma_\mu \gamma_5 N \partial_\nu F^{\nu\mu} \; . 
\end{equation} 
(Here, we have elected to normalize the interaction to the scale of chiral 
symmetry breaking, $\Lambda_\chi=4\pi F_\pi$.) These many-quark anapole 
contributions to 
$\RAp$, which are independent of the electroweak gauge parameter\cite{mh}, 
were first studied in Ref. \cite{khr91,mike} and found in Ref. \cite{mike} 
to carry significant 
theoretical uncertainty. The scale of this uncertainty was estimated in Ref. 
\cite{mike}, and this value was used to obtain the theoretical error in Eq. 
(\ref{eq:gmsep}). (Note that the central value for $\GMS$ given in Eq. 
(\ref{eq:gmsep}) 
is obtained from the experimental asymmetry using the calculation of Ref. 
\cite{mike}). 
 
In order to better constrain the error in $\GMS$ associated with $\RAp$, 
the SAMPLE collaboration performed a second backward angle PV measurement using 
quasielastic (QE) scattering from the deuteron. The asymmetry 
$\alr({\hbox{QE}})$ is significantly less 
sensitive to $\GMS$ than is $\alr(ep)$, but retains a strong 
dependence on $\rateo$, the {\em isovector} part of $\RAp$. The 
calculation of Ref. 
\cite{mike} found the uncertainty in $\RAp$ to be dominated by this 
isovector component---$\rateo\approx -0.34\pm 0.20$---and the goal of 
the deuterium measurement was, therefore, 
to constrain the size of this largest term. A preliminary deuterium 
result was reported at the recent Bates25 Symposium at MIT, and 
suggests that $\rateo$ has the same negative sign as computed in Ref. 
\cite{mike} 
but has considerably larger magnitude, possibly of order unity\cite{bates25}. 
Combining this result with the previous $\alr(ep)$ measurement would 
yield a nearly vanishing value for $\GMS$, rather than the large and 
positive value quoted in Eq. (\ref{eq:gmsep}). 
 
The prospective SAMPLE result for $\rateo$ is remarkable, indicating that 
a higher-order electroweak radiative correction is of the same magnitude 
as, and cancels against, the tree-level amplitude! The occurance of such 
enhanced 
electroweak radiative corrections is rare. Nevertheless, there does exist 
at least one other instance in which higher-order electroweak processes can 
dominate the axial vector hadronic response, namely, the nuclear anapole 
moment. 
The anapole moment of a heavy nucleus grows as $A^{2/3}$ (see, {\em e.g.} 
Refs. \cite{fla84,mh,Hax89} and references therein). Because of the scaling 
with mass 
number, the nuclear AM contribution to a $V(e)\times A({\hbox{nucleus}})$ 
amplitude can 
be considerably larger than the corresponding tree-level $Z^0$-exchange 
amplitude, and 
this $A^{2/3}$ enhancement is consistent with the size of the Cesium AM 
recently determined by the Boulder group using atomic 
parity-violation\cite{apv}. 
The reason 
behind the enhancement of $\rateo$ for the few-nucleon system, however, is 
{\it not} 
understood. The goal of the present paper is to investigate whether there 
exist c 
onventional, hadronic physics effects which can explain the enhancement 
apparently 
implied by the SAMPLE deuterium measurement. 
 
In order to address this question, we revisit the analysis of Ref. \cite{mike}. 
Following Ref. \cite{kaplan}, we re-cast that analysis into the framework 
of heavy baryon chiral perturbation theory (HBChPT) \cite{j1,ijmpe}. 
We carry out a complete calculation of $\rateo$ and $\ratez$ to order 
$1/\lamchis$, including loop diagrams not considered in Refs. 
\cite{mike,kaplan}. We also extend those analyses to include decuplet as 
well as octet intermediate states, magnetic insertions, and SU(3) chiral 
symmetry. 
As in Ref. \cite{mike}, we estimate the 
chiral counterterms at ${\cal O}(1/\lamchis)$ using vector meson saturation. 
However, we go beyond that previous analysis and determine the sign of this 
vector meson contribution phenomenologically. We find that decuplet 
intermediate states and magnetic insertions do not contribute up to the 
chiral order at which we truncate. Also, the effect of SU(3) symmetry, in 
the guise of kaon loops, is generally smaller than the pion loops considered 
previously. In the end, we express 
our results in terms of effective PV hadronic couplings. Some of these 
couplings may be determined 
from nuclear and hadronic PV experiments or detailed calculations (for 
reviews, see Refs. 
\cite{hol89,hh95}), while others are presently 
unconstrained by measurement. Guided by phenomenology and the dimensional 
analysis of Ref. \cite{kaplan}, we estimate the range of possible values 
for the new couplings. We suspect that our estimates are overly generous. 
Nevertheless, we find that -- even under liberal assumptions -- the AM 
contributions to 
$\rateo$ appear unable to enhance the one-quark corrections to the level 
apparently observed 
by the SAMPLE collaboration and, in our conclusions, we speculate on 
possible additional 
sources of enhancement not considered here. 
 
The remainder of the paper is organized as follows. In Section 2, we 
relate the anapole 
couplings $a_{s,v}$ to the radiative corrections, $R_\sst{A}^{T=0,1}$, and 
in Section 3, we 
outline our formalism for computing these couplings in HBChPT. A reader 
already familiar 
with this formalism may wish to skip to Section 4, where we compute the 
chiral loop contributions to the nucleon anapole moment through ${\cal 
O}(1/\lamchis)$. 
We also include the leading $1/\mn$ terms in the heavy baryon expansion, 
which generate 
contributions of ${\cal O}(1/\lamchi\mn)$. 
Section 5 contains the vector meson estimate of the chiral counterterms and 
the determination of the sign, while Section 6 gives our numerical estimate 
of the AM 
contributions to $R_\sst{A}^{T=0,1}$. 
We briefly discuss the phenomenology of hadronic and nuclear PV and what 
that phenomenology may imply about the scale of the unknown low-energy 
constants. 
Section 7 summarizes our conclusions. The Appendices give a detailed 
discussion of 
(A) our formalism, (B) the full set of hadronic PV Lagrangians allowed 
under SU(3) 
symmetry, and (C) graphs, nominally present at ${\cal O}(1/\lamchis)$ but 
whose 
contributions vanish. 
 
\section{Anapole contributions to $R_A$ } 
\label{sec3} 
 
The electron-nucleon parity violating amplitude is generated by the 
diagrams in Figure 2. At tree level this amplitude reads 
\begin{equation} 
iM^\sst{PV} = iM^\sst{PV}_\sst{AV} + iM^\sst{PV}_\sst{VA}\; , 
\end{equation} 
where 
\begin{equation}\label{a} 
iM^\sst{PV}_\sst{AV}= i{G_\mu \over 2\sqrt{2}} l^{\lambda 5} < N |J_\lambda |N> 
\end{equation} 
and 
\begin{eqnarray}\label{b}\nonumber 
iM^\sst{PV}_\sst{VA}&=& i{G_\mu \over 2\sqrt{2}} l^\lambda 
< N |J_{\lambda 5}|N>   \\ 
&=&-i {1-4\sstw \over 2\sqrt{2}} g_A G_\mu \bar e\gamma^\lambda e 
  \bar N \tau_3  \gamma_\lambda \gamma_5 N   \; . 
\end{eqnarray} 
at tree-level in the Standard Model (Figure 2a). 
Here, $J_\lambda$ ($J_{\lambda 5}$) and $ l_\lambda$ ($l_{\lambda 5}$) 
denote the vector (axial vector) weak neutral currents of the quarks and 
electron, 
respectively \cite{MRM94}. The anapole moment interaction of Eq. 
(\ref{eq:anapole}) 
generates additional contributions to $M^\sst{PV}_\sst{VA}$  when a photon is 
exchanged between the nucleon and the electron (Figure 2b). The corresponding 
amplitude is 
\begin{equation} 
\label{eq:manapole} 
iM^\sst{PV}_\sst{AM} = i{(4\pi\alpha)\over\lamchis} {\bar e}\gamma^\lambda 
e{\bar N}(a_s + 
a_v\tau_3) \gamma_\lambda \gamma_5 N \  \ \ \ . 
\end{equation} 
Note that unlike $iM^\sst{PV}_\sst{VA}$, $iM^\sst{PV}_\sst{AM}$ 
contains no $(1-4\sstw)$ suppression. Consequently, 
the relative importance of the anapole interaction is enhanced 
by $1/(1-4\sstw)\sim 10$. This enhancement may be seen explicitly by converting 
Eqs. (\ref{a}) and (\ref{eq:manapole}) into $R_A^\sst{T=0,1}$: 
\begin{equation} 
\label{eq:raanapole} 
\ratez|_{\hbox{anapole}}=-{8\sqrt{2}\pi\alpha\over G_\mu\lamchis} 
{1\over 1-4\sstw}{a_s\over g_A} 
\end{equation} 
\begin{equation} 
\rateo|_{\hbox{anapole}}=-{8\sqrt{2}\pi\alpha\over G_\mu\lamchis} 
{1\over 1-4\sstw}{a_v\over g_A} 
\end{equation} 
 
The constants $a_{s,v}$ contain contributions from loops generated by the 
Lagrangians given in Section 3 and from counterterms in 
the tree-level effective Lagrangian of Eq. 
(\ref{eq:anapole}): 
\begin{equation} 
a_{s,v} = a_{s,v}^\sst{L} + a_{s,v}^\sst{CT}\ \ \ . 
\end{equation} 
In HBChPT, only the parts of the loop amplitudes non-analytic in quark 
masses can be unambigously 
indentified with $a_{s,v}^\sst{L}$. The remaining analytic terms are 
included in $a_{s,v}^\sst{CT}$. In what follows, we compute explicityly 
the various loop 
contributions up through ${\cal O}(1/\lamchis)$, while in principle, 
$a_{s,v}^\sst{CT}$ should be 
determined from experiment. In Section 5, however, 
we discuss a model estimate for 
$a_{s,v}^\sst{CT}$. 
 
Before proceeding with details of the calculation, it is useful to take 
note of the scales 
present in 
Eqs. (\ref{eq:raanapole}). The constants $a_{s,v}$ are generally 
proportional to a product of 
strong and weak meson-baryon couplings. The former are generally of order 
unity, while the size of weak, PV couplings can be expressed in 
terms of $g_\pi=3.8\times 10^{-8}$, the scale of 
charged current contributions\cite{ddh}. One then expects the AM 
contributions to 
the axial radiative corrections to be of order 
\begin{equation} 
\label{eq:rascale} 
R_A^\sst{T=0,1}\sim -{8\sqrt{2}\pi\alpha\over 
G_\mu\lamchis}{1\over 1-4\sstw}{g_\pi\over g_A}\approx -0.01\ \ \ . 
\end{equation} 
In some cases, the PV hadronic couplings may be an order of magnitude 
larger than $g_\pi$. 
Alternatively, chiral singularities arising from loops may also enhance the AM 
effects over the  scale in Eq. (\ref{eq:rascale}). Thus, as we 
show below, the net effect of the AM is anticipated to 
be a 10-20 \% contribution to 
$R_A^\sst{T=0,1}$. 
 
\section{Notations and Conventions} 
\label{sec2} 
 
Since much of the formalism for HBChPT is standard, we relegate a detailed 
summary of 
our conventions to Appendix A. However, some discussion of the effective 
Lagrangians used 
in computing chiral loop contributions to $a_{s,v}$ is necessary here. 
Specifically, we 
require the parity-conserving (PC) and parity-violating (PV) Lagrangians 
involving 
pseudoscalar meson, 
spin-$1/2$ and spin-$3/2$ baryon, and photon fields. For the moment, we 
restrict ourselves to 
SU(2) flavor symmetry and generalize to SU(3) later. The relativistic PC 
Lagrangian for 
$\pi$, $N$, $\Delta$, and $\gamma$ interactions needed here is 
\begin{eqnarray} 
\label{eq:lpc} 
\nonumber 
{\cal L}^\sst{PC}&=&{F_\pi^2\over 4} Tr D^\mu \Sigma D_\mu \Sigma^\dag + 
\bar N (i {\cal D}_\mu \gamma^\mu -m_N) N + g_A \bar N A_\mu \gamma^\mu 
\gamma_5 N\\ \nonumber 
&&+{e\over \Lambda_{\chi}}\bar N (c_s +c_v\tau_3) \sigma^{\mu\nu} 
F^+_{\mu\nu} N \\  \nonumber 
&&-T_i^\mu [(i{\cal D}^{ij}_\alpha\gamma^\alpha -m_\Delta \delta^{ij} 
)g_{\mu\nu}- 
{1\over 4} \gamma_\mu \gamma^\lambda (i{\cal D}^{ij}_\alpha\gamma^\alpha 
-m_\Delta \delta^{ij} ) 
\gamma_\lambda \gamma^\nu \\ \nonumber 
&&+{g_1\over 2} g_{\mu\nu} A_\alpha^{ij} \gamma^\alpha \gamma_5 
+{g_2\over 2} (\gamma_\mu A_\nu^{ij} +A_\mu^{ij} \gamma_\nu )\gamma_5 
+{g_3\over 2} \gamma_\mu A_\alpha^{ij} \gamma^\alpha\gamma_5 \gamma_\nu] 
T_j^\nu \\ \nonumber 
&&+g_{\pi N\Delta} [\bar T^\mu_i (g_{\mu\nu} +z_0 \gamma_\mu\gamma_\nu) 
\omega_i^\nu N 
+\bar N \omega_i^{\nu\dag} (g_{\mu\nu} +z_0 \gamma_\nu\gamma_\mu) T_i^\mu ]\\ 
&&-ie {c_\Delta q_i\over \Lambda_{\chi}}{\bar T}^\mu_i F^+_{\mu\nu}T^\nu_i 
+ [{ ie\over \Lambda_{\chi}}{\bar T}^\mu_3 (d_s +d_v\tau_3) \gamma^\nu\gamma_5 
F^+_{\mu\nu} N +h.c.] 
\end{eqnarray} 
where ${\cal D}_\mu$ is a chiral and electromagnetic (EM) covariant 
derivative, $\Sigma=\exp(i\vec{\tau}\cdot\vec{\pi}/F_\pi)$ is 
the conventional non-linear representation of the pseudoscalar field, 
$N$ is a nucleon isodoublet field, 
$T_\mu^i$ is the 
$\Delta$ field in the isospurion formalism, $F^{\mu\nu}$ is the photon 
field strength tensor, 
and $A_\mu$ is the axial field involving the pseudoscalars 
\begin{equation} 
A_\mu=-{D_\mu\pi\over F_\pi} + {\cal O}(\pi^3)\ \ \ 
\end{equation} 
with $D_\mu$ being the EM covariant derivative. Explicit expressions for 
the fields and the 
transformation properties can be found in Appendix A. The constants $c_s, 
c_v$ determined in terms of the nucleon isoscalar 
and isovector magnetic moments, 
$c_\Delta$ is the $\Delta$ magnetic moment, $d_s, d_v$ are the nucleon 
and delta 
transition magnetic moments, and $z_0$ is the off-shell parameter which is 
not relevant in the 
present work \cite{hhk}. Our convention for $\gamma_5$ is that of Bjorken 
and Drell \cite{BjD}. 
 
In order to obtain proper chiral counting for the nucleon, we 
employ the conventional heavy baryon expansion of 
${\cal L}^\sst{PC}$, and in order to cosistently include the $\Delta$ we 
follow the small scale expansion proposed in 
\cite{hhk}. In this 
approach energy-momenta and the delta and nucleon mass difference $\delta$ are 
both treated 
as ${\cal O}(\epsilon)$ in chiral power counting. The leading order 
vertices in this framework can 
be obtained via $P_+ \Gamma P_+$ where $\Gamma$ is the original 
vertex in the 
relativistic Lagrangian and 
\begin{equation} 
P_{\pm}={1\pm \notv\over 2}\ \ \ . 
\end{equation} 
are projection operators for the large, small components of the Dirac 
wavefunction respectively.  Likewise, the  $O(1/m_N)$ corrections 
are generally propotional to ${P_+ \Gamma P_-/ 
m_N}$. In previous work the parity conserving $\pi N \Delta \gamma$ 
interaction Lagrangians have been obtained to $O({1/ m_N^2})$\cite{hhk}. 
We collect some of the relevant terms below: 
\begin{eqnarray}\nonumber 
{\cal L}^\sst{PC}_v &=& \bar N [iv\cdot D +2g_A S\cdot A]N 
-i {\bar T}^\mu_i [iv\cdot D^{ij} -\delta^{ij} \delta +g_1 S\cdot A^{ij}] 
T_\mu^j \\ \nonumber 
&&+g_{\pi N\Delta}[{\bar T}^\mu_i \omega_\mu^i N 
+ \bar N \omega_\mu^{i\dag} T_i^\mu] \\ \nonumber 
&&+{1\over 2m_N} \bar N \Bigl\{ (v\cdot D)^2-D^2 +[S_\mu, S_\nu][D^\mu, 
D^\nu]\\ 
&&\ \ \ \ -ig_A(S\cdot D v\cdot A +v\cdot A S\cdot D)\Bigr\}N +\cdots 
\end{eqnarray} 
where $S_\mu$ is the Pauli-Lubanski spin operator and  $\delta \equiv m_\Delta 
-m_N$. 
 
The PV analog of Eq. (\ref{eq:lpc}) can be constructed using the chiral fields 
$X^a_{L,R}$ defined in Appendix A and the spacetime transformation 
properties of the 
various fields in Eq. (\ref{eq:lpc}). We find it convenient to follow the 
convention in 
Ref. \cite{kaplan} and separate the PV Lagrangian into its various 
isospin components. The hadronic weak interaction has the form 
\begin{equation}\label{38} 
{\cal H}_\sst{W} = {G_\mu\over\sqrt{2}}J_\lambda J^{\lambda\ \dag}\ + \ 
{\hbox{h.c.}}\ \ \ , 
\end{equation} 
where $J_\lambda$ denotes either a charged or neutral weak current built 
out of quarks. In the Standard Model, the strangeness conserving 
charged currents are pure 
isovector, whereas the 
neutral currents contain both isovector and isoscalar components. 
Consequently, 
${\cal H}_\sst{W}$ contains $\Delta T=0, 1, 2$ pieces and these channels 
must all be accounted for in any realistic hadronic effective theory. 
 
Again for simplicity, we restrict our attention first to the light quark 
SU(2) sector. (A general SU(3) PV meson-baryon Lagrangian is given in the 
Appendix and is 
considerably more complex.) We quote the relativistic Lagrangians, 
but employ 
the heavy baryon projections, as described above, in computing loops. It is 
straightforward to 
obtain the corresponding heavy baryon Lagrangians from those listed below, 
so we do not 
list the PV heavy baryon terms below. For the $\pi N$ sector we have 
\begin{eqnarray}\label{n1} 
{\cal L}^{\pi N}_{\Delta T=0} &=&h^0_V \bar N A_\mu \gamma^\mu N \\ 
\nonumber && \\ 
\label{n2} 
{\cal L}^{\pi N}_{\Delta T=1} &=&{h^1_V\over 2} \bar N  \gamma^\mu N  Tr 
(A_\mu X_+^3) 
-{h^1_A\over 2} \bar N  \gamma^\mu \gamma_5N  Tr (A_\mu X_-^3)\\ 
\nonumber 
&& \ \ \ -{h_{\pi}\over 2\sqrt{2}}F_\pi \bar N X_-^3 N\\ 
\nonumber && \\ 
\label{n3} 
{\cal L}^{\pi N}_{\Delta T=2} &=&h^2_V {\cal I}^{ab} \bar N 
[X_R^a A_\mu X_R^b +X_L^a A_\mu X_L^b]\gamma^\mu N \\ \nonumber 
&& \ \ \ -{h^2_A\over 2} {\cal I}^{ab} \bar N 
[X_R^a A_\mu X_R^b -X_L^a A_\mu X_L^b]\gamma^\mu\gamma_5 N \; . 
\end{eqnarray} 
The above Lagrangian was first given by Kaplan and Savage (KS)\cite{kaplan}. 
However, the coefficients used in our work are 
slightly different from those of Ref. \cite{kaplan} since our definition of 
$A_\mu$ differs by an overall phase (see Appendix A).  Moreover, the 
coefficient of 
the second term in the  original PV $\Delta T =2$ $NN\pi\pi$ Lagrangian 
in Eq. (2.18) was misprinted in the work of KS, and should be 
$2h_A^2$ in their notation instead of $h_V^2$ as given in Eq. (2.18) of 
\cite{kaplan}. 
 
The term proportional to $h_\pi$ contains no derivatives and, 
at leading-order in $1/F_\pi$, 
yields the PV $NN\pi$ Yukawa coupling traditionally used in meson-exchange 
models for the 
PV NN interaction \cite{ddh,haxton}. The PV $\gamma$-decay of $^{18}$F can 
be used to 
constrain the value of $h_\pi$ in a nuclear model-independent way as 
discussed in Ref. 
\cite{haxton}, resulting in $h_\pi=(0.7\pm 2.2)g_\pi$ \cite{hh95}. Future 
PV experiments are planned using 
light nuclei to confirm the $^{18}$F result. The coupling $h_\pi$ has also 
received considerable 
theoretical attention\cite{ddh,fcdh,t1,t2} and is particularly interesting 
since it receives no 
charged current contributions at leading order. 
 
Unlike the PV Yukawa interaction, the vector and axial vector terms in Eqs. 
(\ref{n1}-\ref{n3}) 
contain derivative  interactions. The terms containing $h_A^1, h_A^2$ start 
off with $NN\pi\pi$ 
interactions, while all the other terms start off as $NN\pi$. 
Such derivative interactions have not been included in conventional analyses 
of nuclear and hadronic PV experiments. Consequently, the experimental 
constraints on the 
low-energy constants $h_V^i$, $h_A^i$ are unknown. The authors of Ref. 
\cite{kaplan} used 
simple dimensional arguments and factorization limits to estimate their 
values, and we present 
additional phenomenological considerations in Section 6 below. We 
emphasize, however, that 
the present lack of knowledge of these couplings introduces additional 
uncertainties into 
$R_A^{T=0,1}$. 
 
In addition to purely hadronic PV interactions, one may also write down PV 
EM interactions 
involving baryons and mesons\footnote{Note that the hadronic derivative 
interactions of 
Eqs. (\ref{n1}-\ref{n3}) also contain $\gamma$ fields as required by 
gauge-invariance}. 
The anapole interaction of Eq. (\ref{eq:anapole}) represents 
one such interaction, arising at ${\cal O}(1/\lamchis)$ and involving no 
$\pi$'s. There also 
exist terms at ${\cal O}(1/\lamchi)$ which include at least one $\pi$: 
\begin{equation}\label{n4} 
{\cal L}^{\gamma N}\sst{PV} ={c_1\over \Lambda_{\chi}} \bar N 
\sigma^{\mu\nu} [F^+_{\mu\nu}, 
X_-^3]_+ N  +{c_2\over \Lambda_{\chi}} \bar N \sigma^{\mu\nu} F^-_{\mu\nu} N 
+{c_3\over \Lambda_{\chi}} \bar N \sigma^{\mu\nu} [F^-_{\mu\nu}, X_+^3]_+ N 
\; . 
\end{equation} 
 
The corresponding PV Lagrangians involving a $N\to\Delta$ transition are 
somewhat more 
complicated. The analogues of Eqs. (\ref{n1}-\ref{n3}) are 
\begin{eqnarray}\label{d1}\nonumber 
{\cal L}^{\pi\Delta N}_{\Delta I=0} &=&f_1 \epsilon^{abc} \bar N i\gamma_5 
[X_L^a A_\mu X_L^b +X_R^a A_\mu X_R^b] T_c^\mu \\ 
&& +g_1 \bar N [A_\mu, X_-^a]_+ T^\mu_a+g_2 \bar N [A_\mu, X_-^a]_+ T^\mu_a 
+{\hbox{h.c.}}\\ 
\nonumber && \\ 
\label{d2} \nonumber 
{\cal L}^{\pi\Delta N}_{\Delta I=1} &=& 
f_2 \epsilon^{ab3} \bar N i\gamma_5 [A_\mu, X_+^a]_+ T^\mu_b 
+f_3 \epsilon^{ab3}\bar N i\gamma_5[A_\mu, X_+^a]_- T^\mu_b \\ \nonumber 
&&+g_3\bar N [(X_L^a A_\mu X_L^3-X_L^3 A_\mu X_L^a)- 
(X_R^a A_\mu X_R^3-X_R^3 A_\mu X_R^a)] T^\mu_a\\ \nonumber 
&&+g_4 \{\bar N [3X_L^3 A^\mu (X_L^1 T^1_\mu +X_L^2 T^2_\mu ) + 3(X_L^1 A^\mu 
X_L^3 T^1_\mu 
+X^2_L A^\mu X^3_L T^2_\mu) \\ 
&&-2 (X_L^1 A^\mu X_L^1 +X_L^2 A^\mu X_L^2-2X_L^3 A^\mu X_L^3)T^3_\mu] 
-(L\leftrightarrow R) \} 
+{\hbox{h.c.}} \\ 
\nonumber && \\ 
\label{d3} \nonumber 
{\cal L}^{\pi\Delta N}_{\Delta I=2} &=& 
f_4 \epsilon^{abd} {\cal I}^{cd}\bar N i\gamma_5  [X_L^a A_\mu X_L^b 
+X_R^a A_\mu X_R^b]T^\mu_c \\ \nonumber 
&&+f_5 \epsilon^{ab3} \bar N i\gamma_5  [X_L^a A_\mu X_L^3+X_L^3 A_\mu X_L^a 
+(L\leftrightarrow R)]T^\mu_b \\ 
&&+g_5 {\cal I}^{ab}\bar N [A_\mu, X_-^a]_+ T^\mu_b 
+g_6 {\cal I}^{ab}\bar N [A_\mu, X_-^a]_+ T^\mu_b 
+{\hbox{h.c.}}\ \ \ , 
\end{eqnarray} 
where the terms containing $f_i$ and $g_i$ start off with single and two pion 
vertices, respectively. 
 
Finally, we consider PV $\gamma \Delta N$ interactions: 
\begin{eqnarray}\label{d4} 
{\cal L}^{\gamma \Delta N}_\sst{PV} &=&ie{d_1\over \Lambda_{\chi}} {\bar 
T}^\mu_3 \gamma^\nu 
F^+_{\mu\nu}N +ie{d_2\over \Lambda_{\chi}} {\bar T}^\mu_3 \gamma^\nu 
[F^+_{\mu\nu}, X_+^3]_+ N \\ \nonumber 
&&+ie{d_3\over \Lambda_{\chi}} {\bar T}^\mu_3 \gamma^\nu 
[F^+_{\mu\nu},X_+^3]_- N 
+ie{d_4\over \Lambda_{\chi}} {\bar T}^\mu_3 \gamma^\nu\gamma_5F^-_{\mu\nu}N 
\\ \nonumber 
&&+ie{d_5\over \Lambda_{\chi}} {\bar T}^\mu_3 \gamma^\nu\gamma_5 
[F^+_{\mu\nu}, X_-^3]_+ N 
+ie{d_6\over \Lambda_{\chi}} {\bar T}^\mu_3 \gamma^\nu\gamma_5 
[F^-_{\mu\nu}, X_+^3]_+ N \\ 
&& +ie{d_{7}\over \Lambda_{\chi}} {\bar T}^\mu_3 \gamma^\nu [F^-_{\mu\nu}, 
X_-^3]_+ N 
+ie{d_{8}\over \Lambda_{\chi}} {\bar T}^\mu_3 \gamma^\nu [F^-_{\mu\nu}, 
X_-^3]_- N + {\hbox{h.c.}}  . 
\end{eqnarray} 
The PV  $\gamma \Delta N$ vertices $d_{1-3}$, $d_{4-6}$ and $d_{7-8}$ 
are associated at leading order in $1/F_\pi$ with zero, one and two pion 
vertices, respectively. All the vertices in 
(\ref{n1})-(\ref{d4})  are ${\cal O}(p)$ or ${\cal O}({1/ \Lambda_{\chi}})$ 
except 
$h_{\pi }$, which is Yukawa interaction and of $O(p^0)$. As we discuss in 
Appendix C, we 
do not require PV interactions involving two $\Delta$ fields. 
 
\section{Chiral Loops} 
\label{sec:chiral} 
 
The contributions to $a_{s,v}$ arising from the Lagrangians of Eqs 
. 
(\ref{n1}-\ref{n3}) 
are shown in Figure 3. We regulate the associated integrals using dimensional 
regularization 
(DR) and absorb the divergent---$1/(d-4)$---terms into the counterterms, 
$a_{s,v}^\sst{CT}$. 
The leading contributions arise from the PV Yukawa coupling $h_\pi$ 
contained in 
the loops of 3a-f. To ${\cal O}(1/\lamchis)$, the diagrams 3e,f containing a 
photon insertion on a nucleon line do not contribute. The reason is readily 
apparent 
from examination of the integral associated with the amplitude of Figure 3e: 
\begin{eqnarray} \label{eq:integral1} \nonumber 
iM_{3e} =ie_N h_{\pi} v\cdot\varepsilon {\sqrt{2} g_A\over F_\pi} \int 
{d^Dk\over (2\pi)^D} 
{i(S\cdot k)\over v\cdot k} {i\over v\cdot (q+k)}{i\over k^2-m_\pi^2 
+i\epsilon}    & \\ 
=-ie_N h_{\pi} v\cdot\varepsilon {2\sqrt{2} g_A\over F_\pi} S_\mu 
\int_0^\infty {sds}\ \int_0^1 du\  \int {d^Dk\over (2\pi)^D} 
{k_\mu\over [k^2+s v\cdot k 
+us v\cdot q +\mpis]^3} \  ,& 
\end{eqnarray} 
where $q_\mu$ is the photon momentum, $\varepsilon$ is the photon 
polarization vector, $s$ has the dimensions of mass, 
and we have Wick rotated to Euclidean momenta in the second 
line. From this form it is clear that $iM_{3e}\propto S\cdot v = 0$. 
The sum of the non-vanishing diagrams Figure 3a-d yields a gauge invariant 
leading order result, which is 
purely isoscalar: 
\begin{equation}\label{leading} 
a_s^{\sst{L}}(Y1)= -{\sqrt{2}\over 24}e g_A h_{\pi }{\Lambda_\chi\over 
m_\pi } \; . 
\end{equation} 
 
As the PV Yukawa interaction is of order $O(p^0)$, we 
need to consider higher order corrections involving this interaction, which 
arise from 
the $1/m_N$ expansion of the nucleon propagator and various vertices. Since 
$P_+ \cdot 1\cdot 
P_-=0$, there is no $1/m_N $ correction to the PV Yukawa vertex. From the 
$1/m_N$ 
${\bar N} N$ terms in Eq. (\ref{eq:lpc}) we have 
\begin{equation}\label{prop} 
a_s^{\sst{L}}(Y2)= {7\sqrt{2}\over 48\pi}e g_A h_{\pi }{\lamchi\over m_N} 
\ln ({\mu\over m_\pi})^2\; , 
\end{equation} 
where $\mu$ is the subtraction scale introduced by DR. 
Finally, the $1/m_N$ correction to the strong $\pi NN$ vertex, contained in 
the term 
$\propto g_A$ in 
Eq. (\ref{eq:lpc}), yields 
\begin{equation}\label{vertex} 
a_s^{\sst{L}}(Y3)=- {\sqrt{2}\over 48\pi}e g_A h_{\pi}{\Lambda_\chi\over 
m_N}  \ln ({\mu\over m_\pi})^2\; . 
\end{equation} 
These terms are also isoscalar, and the results in Eqs. 
(\ref{leading}-\ref{vertex}) are fully contained 
in the previous analyses of Refs. \cite{mike,mh,kaplan}. 
 
For the interactions in Eqs. (\ref{n1}-\ref{n3}) containing $h_V^i$, the 
eight diagrams Figure 3a-h must be considered. Their 
contribution is  purely isovector--- 
\begin{equation}\label{vector} 
a_v^{\sst{L}}(V)={1\over 6}eg_A (h_V^0+{4\over 3}h_V^2) 
\ln ({\mu\over m_\pi})^2\; . 
\end{equation} 
---and was not included in previous analyses. 
 
The contribution generated from the two-pion PV axial vertices in Eqs. 
(\ref{n2}-\ref{n3}) 
comes only from the loop Figure 3i and contains both isovector and 
isoscalar components: 
\begin{equation}\label{axia} 
a_s^{\sst{L}}(A)+a_v^{\sst{L}}(A)\tau_3=-{1\over 3}e(h_A^1+h_A^2\tau_3) 
\ln ({\mu\over m_\pi})^2\; . 
\end{equation} 
a result first computed in Ref. \cite{kaplan}. 
 
In principle, a variety of additional contributions will arise at ${\cal 
O}(1/\lamchis)$. 
For example, insertion of the nucleon magnetic moments ({\it i.e.} 
the terms in Eq. (\ref{eq:lpc}) 
containing $c_{s,v}$) into the loops Figure 3e,f---resulting in the 
loops of Figure 5a,b---would in principle generate terms of 
${\cal O}(1/\lamchis)$ 
when the PV Yukawa interaction is considered. As shown in Appendix C, 
however, such contributions vanish at this order. Similarly, 
the entire set of $\Delta$ intermediate 
state contributions shown 
in Figure 4, as well as those generated by ${\cal L}^{\gamma N}_\sst{PV}$ and 
${\cal L}^{\gamma\Delta 
N}_\sst{PV}$ in Figure 6, vanish up to ${\cal O}(1/\lamchis)$. The reasons 
for the vanishing of 
these various possible contributions is discussed in Appendix C. Thus, the 
complete set of 
SU(2) loop contributions up to ${\cal O}(1/\lamchis)$ are given in Eqs. 
(\ref{leading}-\ref{axia}). 
 
Because $m_c-m_s >> m_s-m_{u,d}$ and $\lamchi >> m_s$, it may be appropriate 
to treat the lightest strange and non-strange 
hadrons on a similar footing and extend the foregoing discussion to SU(3) 
chiral symmetry. A 
similar philosophy has been adopted by several authors in studying the 
axial charges and magnetic 
moments of the lightest baryons \cite{j1,k1,k2,k3,k4}. In what follows, we 
consider the possibility 
that kaon loop contributions, introduced by the consideration of SU(3) 
symmetry, may 
further enhance the anapole contribution to $R_A$. 
 
Before proceeding along these lines, however, 
one must raise an important caveat. When 
kaon loop corrections are included in a HBChPT analysis, higher order 
chiral corrections may go as powers of $\mk/\lamchi\sim 0.5$. Consequently, 
the convergence of the SU(3) chiral expansion 
remains a subject of debate\cite{convergence}. Fortunately, no such 
factors appear in the 
present analysis through ${\cal O}(1/\lamchis)$ so that at this order, 
we find that kaon 
loop effects in $R_A$ are generally tiny compared to those involving pion 
loops. 
Whether or not higher-order terms ({\em e.g.}, those 
of ${\cal O}(1/\lamchis\times\mk/\lamchi)$ contribute as strongly as those 
considered here remains a separate, open question. 
 
To set our notation, we give the leading strong-interaction SU(3) 
Lagrangian. Since the 
$K^0$ and $\eta$ are neutral, loops containing these mesons do not 
contribute to the AM 
through ${\cal O}(1/\lamchis)$ and we do not include their strong couplings 
below. 
For the proton the 
possible intermediate states are $\Sigma^0 K^+, \Lambda K^+$ while for the 
neutron only $\Sigma^- 
K^+$ can appear. The necessary vertices derive from 
\begin{eqnarray} 
{\cal L}&=&2g_A\bar N S\cdot A N +2g_{N\Lambda K}[(\bar N S\cdot K)\Lambda 
+\bar\Lambda (S\cdot K^\dag 
N)]\\ 
\nonumber 
&& +2g_{N\Sigma K} [S\cdot K^\dag \bar\Sigma N +\bar N \Sigma S\cdot K] \; , 
\end{eqnarray} 
where $g_{N\Lambda K}=-[(1+2\alpha)/ \sqrt{6}]g_A$, $g_{N\Sigma 
K}=(1-2\alpha )g_A$ with $g_A =D+F$, $\alpha=F/(D+F)$ and $D,F$ are the 
usual $SU(3)$ symmetric and 
antisymmetric coupling constants. 
 
The general pesudoscalar octet and baryon octet PV Lagrangians are given in 
the Appendix B. They contain four independent PV Yukawa couplings, 20 axial 
vector couplings ($h_A$-type), and 22 vector couplings ($h_V$-type). 
For simplicity, we combine the SU(3) 
couplings into combinations specific to various hadrons---{\it e.g.} the 
leading PV Yukawa interactions are 
\begin{eqnarray}\label{yu}\nonumber 
{\cal L}^{1\pi}_{\hbox{Yukawa}} =-ih_{\pi} (\bar p n \pi^+ -\bar n p \pi^-) 
-ih_{p\Sigma^0 K}(\bar p \Sigma^0 K^+-{\bar{\Sigma^0}} p K^-) &\\ 
-ih_{n\Sigma^- K}(\bar n \Sigma^- K^+-{\bar{\Sigma}^-} n K^-) 
-ih_{p\Lambda K}(\bar p \Lambda K^+-{\bar\Lambda} p K^-) 
+\cdots &\; . 
\end{eqnarray} 
In terms of the SU(3) couplings listed in Appendix B, the $h_{BBM}$ have the 
form 
\begin{eqnarray}\nonumber 
h_{\pi} &=& -2\sqrt{2} (h_1 +h_2) \\ \nonumber 
h_{p\Sigma^0 K} &=& -[h_1-h_2+\sqrt{3}(h_3-h_4)] \\ \nonumber 
h_{n\Sigma^- K} &=& \sqrt{2} h_{p\Sigma^0 K} \\ 
h_{p\Lambda K} & = & \left[ {h_1\over\sqrt{3}} + \sqrt{3} h_2 + h_3 + 3 
h_4\right] \ \ \ . 
\end{eqnarray} 
 
Similarly, we write for the vector PV interaction 
\begin{eqnarray}\label{ve}\nonumber 
{\cal L}^{1\pi}_{V} &=&-{h_V^{pn\pi^+}\over \sqrt{2}F_\pi} \bar p\gamma^\mu 
n D_\mu\pi^+ 
-{h_V^{p\Sigma^0 K^+}\over \sqrt{2}F_\pi} \bar p \gamma^\mu\Sigma^0 D_\mu 
K^+ \\ 
&&-{h_V^{n\Sigma^- K^+}\over \sqrt{2}F_\pi} \bar n \gamma^\mu\Sigma^- D_\mu K^+ 
-{h_V^{p\Lambda K^+}\over \sqrt{2}F_\pi} \bar p \gamma^\mu\Lambda D_\mu K^+ 
+ h.c. +\cdots \; , 
\end{eqnarray} 
and for the axial PV two pion and kaon interactions 
\begin{eqnarray}\label{ax}\nonumber 
{\cal L}^{2\pi}_{A} &=& i{h_A^{p\pi}\over F^2_\pi}\bar p\gamma^\mu \gamma_5 p 
(\pi^+ D_\mu \pi^--\pi^- D_\mu \pi^+) 
+i{h_A^{pK}\over F^2_\pi}\bar p\gamma^\mu \gamma_5 p 
(K^+ D_\mu K^--K^- D_\mu K^+) \\ 
\nonumber 
&&+i{h_A^{n\pi}\over F^2_\pi}\bar n\gamma^\mu \gamma_5 n 
(\pi^+ D_\mu \pi^--\pi^- D_\mu \pi^+) 
+i{h_A^{nK}\over F^2_\pi}\bar n\gamma^\mu \gamma_5 n 
(K^+ D_\mu K^--K^- D_\mu K^+) \\ 
&&+\cdots \; . 
\end{eqnarray} 
Expressions for these PV vector and axial coupling constants in terms of SU(3) 
constants appear in Appendix B. For illustrative purposes, it is useful to 
express the nucleon-pion couplings in terms of the $h_{V,A}^i$ of Eqs. 
(\ref{n1}-\ref{n3}) for the SU(2) sector: 
\begin{eqnarray} 
\label{eq:su2lim} \nonumber 
h_V^{pn\pi^+} & = & h_V^0+{4\over 3} h_V^2 \\ \nonumber 
h_A^{p\pi} & = & h_A^1 + h_A^2 \\ 
h_A^{n\pi} & = & h_A^1 - h_A^2\ \ \ . 
\end{eqnarray} 
 
The leading order contributions to $a_{s,v}$ arise only from the loops in 
Figure 3 where 
a photon couples to a charged meson. The charged kaon loop contributions to 
the $a_{s,v}$ can be 
obtained from the corresponding formulae for the $\pi$-loop 
terms by making 
simple replacements 
of couplings and masses. For example, for 
the PV Yukawa interactions, these replacements are: (a) for the proton case, 
$m_\pi \to m_K, h_{\pi } \to h_{p\Sigma^0 K}$, $g_A \to g_{N\Sigma^0 
K^+}={g_{N\Sigma K}/ 
\sqrt{2}}$ for $\Sigma^0 K^+$ intermediate states and $h_{\pi} \to 
h_{p\Lambda K}$, $g_A \to g_{N\Lambda K}$ for 
$\Lambda K^+$; (b) for the neutron case,  $h_{\pi} \to h_{n\Sigma^- K}$, $g_A 
\to g_{N\Sigma^- K^+}= 
g_{N\Sigma K}$ for $\Sigma^- K^+$ intermediate state for the neutron case. 
Similar replacements hold for the vector PV coupling contributions. 
For the axial PV 
two-pion contribution we need only make the 
replacement $h_A^{p\pi}\to h^{pK}_A, 
m_\pi\to m_K$. 
 
Upon making these substitutions, we obtain the complete heavy baryon loop 
contribution to 
${\cal O}(1/\lamchis)$ in SU(3): 
\begin{eqnarray}\label{rs}\nonumber 
a_s^\sst{L}&=&{\sqrt{2}\over 24}  g_A h_{\pi } 
[-{\Lambda_{\chi}\over m_\pi} +{3\over \pi} {\Lambda_{\chi}\over m_N}\ln 
({\mu\over m_\pi})^2] 
 \\ \nonumber 
&&-{\sqrt{3}\over 144}  (1+2\alpha )g_A h_{p\Lambda K} 
[-{\Lambda_{\chi}\over m_K} +{3\over \pi} {\Lambda_{\chi}\over m_N}\ln 
({\mu\over m_K})^2] 
\\ \nonumber 
&& +{\sqrt{2}\over 32}  (1-2\alpha ) g_A h_{n\Sigma^- K} 
[-{\Lambda_{\chi}\over m_K} +{3\over \pi} {\Lambda_{\chi}\over m_N}\ln 
({\mu\over m_K})^2] 
 \\ \nonumber 
&&-{1\over 6}  (h^{p\pi}_A+h^{n\pi}_A) 
\ln ({\mu\over m_\pi})^2 
-{1\over 6}  (h^{pK}_A+h^{nK}_A) 
\ln ({\mu\over m_K})^2  \\ \nonumber 
&&+{1\over 12}  (1-2\alpha ) g_A (h_V^{n\Sigma^- K^+} +{h_V^{p\Sigma^0 
K^+}\over \sqrt{2}}) 
\ln ({\mu\over m_K})^2 \\ 
&&-{\sqrt{6}\over 72}  (1+2\alpha ) g_A h_V^{p\Lambda K^+} \ln ({\mu\over 
m_K})^2  \\ 
\label{rv}\nonumber 
a_v^\sst{L}&=&-{\sqrt{2}\over 96}  (1-2\alpha ) g_A h_{n\Sigma^- K} 
[-{\Lambda_{\chi}\over m_K} +{3\over \pi} {\Lambda_{\chi}\over m_N}\ln 
({\mu\over m_K})^2] 
 \\ \nonumber 
&&-{\sqrt{3}\over 144}  (1+2\alpha )g_A h_{p\Lambda K} 
[-{\Lambda_{\chi}\over m_K} +{3\over \pi} {\Lambda_{\chi}\over m_N}\ln 
({\mu\over m_K})^2] 
\\  \nonumber 
&&-{1\over 6}  (h^{p\pi}_A-h^{n\pi}_A) 
\ln ({\mu\over m_\pi})^2 
-{1\over 6}  (h^{pK}_A-h^{nK}_A) 
\ln ({\mu\over m_K})^2  \\ \nonumber 
&&-{1\over 6}   g_A h_V^{pn \pi^+} \ln ({\mu\over m_\pi})^2 \\ \nonumber 
&&+{1\over 12}  (1-2\alpha ) g_A (-h_V^{n\Sigma^- K^+} +{h_V^{p\Sigma^0 
K^+}\over \sqrt{2}}) 
\ln ({\mu\over m_K})^2  \\ 
&&-{\sqrt{6}\over 72}  (1+2\alpha ) g_A h_V^{p\Lambda K^+} \ln ({\mu\over 
m_K})^2  \; . 
\end{eqnarray} 
 
\section{Low-energy Constants and Vector Mesons} 
\label{sec4} 
 
A pure ChPT treatment of the anapole contributions to $R_A$ would use a 
measurment of the axial term in $\alr(ep)$ and $\alr(QE)$, together with the 
non-analytic, long-distance loop contributions, $a_{s,v}^\sst{L}$, to 
determine 
the low-energy constants, $a_{s,v}^\sst{CT}$. In the present case, however, 
we wish to determine whether there exist reasonable hadronic mechanisms 
which can 
enhance the low-energy constants to the level suggested by the SAMPLE 
results. Thus, 
we attempt to estimate $a_{s,v}^\sst{CT}$ theoretically. 
 
Because they are governed in part by the short-distance ($r> 1/\lamchi$) 
strong interaction, 
$a_{s,v}^\sst{CT}$ are difficult to compute from first principles in QCD. 
Nevertheless, 
experience with ChPT in the pseudscalar meson sector and with the 
phenomenology of 
nucleon EM form factors suggests a reasonable model approach. 
It is well known, for example, that in the ${\cal O}(p^4)$ 
chiral Lagrangian describing pseudoscalar interactions, the low-energy 
constants are well-described by the exchange of heavy mesons\cite{egpr}. 
In particular, the 
charge radius of the pion receives roughly a 7\% long-distance loop 
contribution, while the 
remaining 93\% is saturated by $t$-channel exchange of the $\rho^0$. 
Similarly, in the baryon sector, dispersion relation analyses of the isovector 
and isoscalar nucleon electromagnetic form factors indicate important 
contributions from the lightest vector mesons\cite{Hoh76}. Thus, it seems 
reasonable to assume that 
$t$-channel exchange of vector mesons also plays an important role in the 
short-distance physics 
associated with the anapole moment. 
 
With these observations in mind, we estimate the coefficients 
$a_{s,v}^\sst{CT}$ in 
the approximation that they are saturated by $t$-channel exchange of the 
lightest 
vector mesons, as shown in Figure 7. 
Here parity-violation enters through the 
vector meson-nucleon interaction vertices. 
We also use a similar picture for the electromagnetic nucleon form factors 
to determine 
the overall phase of $a_{s,v}^\sst{CT}$ in the vector meson dominance 
approximation. To that 
end we require the PC and PV vector meson-nucleon Lagrangians \cite{ddh}: 
\begin{eqnarray} 
{\cal L}^{PC}_{\rho NN}& =& g_{\rho NN} \bar N 
[\gamma_\mu +\kappa_\rho {i\sigma_{\mu\nu} q^\nu\over 2m_N}] 
{\tau \cdot \rho^\mu} N \\ 
{\cal L}^{PC}_{\omega NN} &=&g_{\omega NN} \bar N 
[\gamma_\mu +\kappa_\omega {i\sigma_{\mu\nu} q^\nu\over 2m_N}] 
 \omega^\mu N \\ 
{\cal L}^{PC}_{\phi NN} &=&g_{\phi NN} \bar N 
[\gamma_\mu +\kappa_\phi {i\sigma_{\mu\nu} q^\nu\over 2m_N}] 
 \phi^\mu N 
\end{eqnarray} 
and 
\begin{eqnarray} 
{\cal L}^{PV}_{\rho NN} &=& \bar N 
\gamma^\mu \gamma_5 \rho^0_\mu [ h^1_{\rho } + 
(h^0_{\rho } +{h^2_{\rho }\over \sqrt{6}} )\tau_3] N  \;  \\ 
{\cal L}^{PV}_{\omega NN} &=& \bar N 
\gamma^\mu \gamma_5 \omega_\mu [ h^0_{\omega } +h^1_{\omega }\tau_3] N  \; \\ 
{\cal L}^{PV}_{\phi NN} &=& \bar N 
\gamma^\mu \gamma_5 \phi_\mu [ h^0_{\phi } +h^1_{\phi }\tau_3] N  \; . 
\end{eqnarray} 
(Note that we have adopted a different convention for $\gamma_5$ than used in 
Ref. \cite{ddh}.) The coupling constants $h^i_{\rho,\omega,\phi}$ were 
estimated in Refs. 
\cite{ddh,fcdh} and have also been constrained by a variety of hadronic 
and nuclear 
parity-violating experiments (for a review, see Ref. \cite{haxton}). 
 
For the $V-\gamma$ transition amplitude, we use 
\begin{equation} 
{\cal L}_{V\gamma} = {e\over 2f_V} F^{\mu\nu}V_{\mu\nu}  \; , 
\end{equation} 
where $e$ is the charge unit, $f_V$ is the $\gamma$-$V$ conversion constant 
($V=\rho^0,\omega,\phi$), and $V_{\mu\nu}$ is the corresponding vector meson 
field tensor. (This 
gauge-invariant Lagrangian ensures that the diagrams of Figure 7 do not 
contribute to the charge of 
the nucleon.) The amplitude of Figure 7 then becomes 
\begin{equation}\label{as} 
a_s^\sst{CT}(VMD) ={ h^1_{\rho }\over f_\rho}({\Lambda_\chi\over m_\rho})^2 
+{ h^0_{\omega }\over f_\omega}({\Lambda_\chi\over m_\omega})^2 
+{ h^0_{\phi }\over f_\phi}({\Lambda_\chi\over m_\phi})^2  \; , 
\end{equation} 
\begin{equation}\label{av} 
a_v^\sst{CT}(VMD) ={ h^0_{\rho }+{h^2_{\rho }/ \sqrt{6}}\over 
f_\rho}({\Lambda_\chi\over 
m_\rho})^2  +{ h^1_{\omega }\over f_\omega}({\Lambda_\chi\over m_\omega})^2 
+{ h^1_{\phi }\over f_\phi}({\Lambda_\chi\over m_\phi})^2 \; . 
\end{equation} 
The parity violating rho-pole contribution was first derived in 
\cite{mike,mh}. However, the 
relative sign  between $h_{\rho N}^i$ and $f_\rho$ is undetermined from the 
diagram of Figure 7 alone.  Nevertheless, we can 
fix the overall phase using two phenomenological inputs. 
Parity violating experiments in the p-p system constrain the 
sign of the combination 
$g_{\rho N} h_{\rho N}^i$ \cite{bhj,des,haxton}. In particular, the scale 
of the longitudinal 
analyzing power, $A_L$, is set by the combination of constants 
\begin{equation} 
\label{eq:pplong} 
A_L \propto g_{\rho NN} (2+\kappa_V)[h_\rho^0+h_\rho^1 +h_\rho^2/\sqrt{6}] + 
   g_{\omega NN} (2+\kappa_S) [h_\omega^0+ h_\omega^1] \ \ \ , 
\end{equation} 
where the constant of proportionality is positive, $\kappa_V=3.7$ and 
$\kappa_S=-0.12$. 
Using the standard values for the strong $VNN$ couplings, one finds that 
$A_L$ has roughly 
the same sensitivity to each of the $h_V^i$ appearing in Eq. 
(\ref{eq:pplong}) (modulo the 
$1/\sqrt{6}$ coefficient of $h_\rho^2$). From the 45 MeV experiment 
performed at SIN \cite{Kis87}, 
for example, one obtains the approximate constraint \cite{hh95} 
\begin{equation} 
\label{eq:ppcons} 
h_\rho^0+h_\rho^1+h_\rho^2/\sqrt{6}+h_\omega^0+h_\omega^1\sim -28\pm 4 
\ \ \ , 
\end{equation} 
where the $h_V^i$ have are expressed in units of $g_\pi$ and where a 
positive sign has 
been assumed for $g_{VNN}$. Given this constraint, it is very unlikely 
that the product 
$(h_\rho^0+h_\rho^2/\sqrt{6}) g_{\rho NN} > 0$ unless the corresponding 
products involving $h_\rho^1$ and $h_\omega^{0,1}$ in Eq. 
(\ref{eq:pplong}) obtain anomalously large, negative values. In fact, a fit 
to hadronic 
and nuclear PV observables in Ref. \cite{haxton} strongly favors a phase 
difference between 
the strong and weak $VNN$ couplings. 
 
Experimentally, one also knows the  isovector nucleon charge radius 
\begin{equation}\label{exp} 
\langle r^2\rangle^{T=1}_\sst{EXP} = 6 
{dF_1(q^2)\over dq^2}|_{q^2=0} > 0 
\; , 
\end{equation} 
where 
\begin{equation} 
<p' | j_\mu^{T=1}(0)| p> = e\bar u(p' ) [F_1 (q^2) +{i\sigma_{\mu\nu} 
q^\nu\over 2m_N}F_2 (q^2)] 
u(p)\ \ \ . 
\end{equation} 
 
One may reasonably approximate the $\rho^0$ contribution to $\langle 
r^2\rangle^{T=1}$ 
using VMD \cite{Hoh76}. The calculation is the same as above but with 
the weak hadronic coupling replaced by the strong coupling. The result is 
\begin{equation} 
F_1^{\rho^0} (q^2)={g_{\rho NN}\over f_{\rho}} {q^2\over q^2 -m_\rho^2}\ \ \ . 
\end{equation} 
Then we have 
\begin{equation}\label{vmd2} 
{dF^{VMD}_1(q^2)\over dq^2}|_{q^2=0}=- {g_{\rho NN}\over f_{\rho}m_\rho^2} 
\; . 
\end{equation} 
Comparing Eqs. (\ref{exp}) and (\ref{vmd2}), and noting that the $\rho^0$ 
generates a positive contribution to $\langle r^2\rangle^{T=1}$ 
\cite{Hoh76}, we arrive 
at ${g_{\rho NN}/ f_\rho}<0$.  Combining this result with $g_{\rho NN} 
h^i_{\rho } <0$ as 
favored by the ${\vec p} p$ experiments 
\cite{bhj,des,haxton} we obtain the relative sign between $h_{\rho }^i$ and 
$f_\rho$: $h^i_{\rho 
}/ f_\rho>0$. Accordingly we determine the relative signs for PV 
$\omega, \phi$-nucleon 
coupling constants. 
 
\section{The scale of $R_A$} 
\label{sec5} 
 
Expressions for the anapole contributions to $\ratez$ and $\rateo$ in terms 
of the 
$a_{s,v}$ appear in Eq. (\ref{eq:raanapole}). We may now 
use these expressions, 
along 
with the results in Eqs. (\ref{rs}-\ref{rv}) and (\ref{as}-\ref{av}), to obtain 
a numerical estimate for the $R_\sst{A}^{T=0,1}|_{\hbox{anapole}}$. 
To do so,  we use the global fit value for the weak mixing angle in the 
on-shell scheme, $\sstw =0.2230$ \cite{pdg}, $g_A=1.267\pm0.004$ \cite{pdg}, 
$f_\rho=5.26$ \cite{sakurai}, $f_\omega =17, f_\phi=13$ \cite{hm}, 
$\alpha=F/(D+F)=0.36$, 
$\mu=\Lambda_{\chi} $. We express all the PV coupling 
constants in units of $g_\pi =3.8 \times 10^{-8}$ as is traditionally done 
\cite{fcdh,ddh}. We obtain 
\begin{eqnarray}\label{RAS}\nonumber 
\ratez|_{\hbox{anapole}}&=&10^{-2}\{ 0.17 h_{\pi} +h_A^1 
-0.0036 h_{n\Sigma^- K}  \\ \nonumber 
&&-0.033 (h_V^{n\Sigma^- K^+} 
 +{h_V^{p\Sigma^0 K^+}\over \sqrt{2}}) 
+0.2  (h^{pK}_A+h^{nK}_A) -0.006 h_{p\Lambda K} \\ 
&&+0.088 h_V^{p\Lambda K^+} -0.26 |h^1_{\rho }| -0.08  |h^0_{\omega 
}|-0.05 |h^0_{\phi }| 
\}  \\ 
\label{RAV}\nonumber 
\rateo|_{\hbox{anapole}}&=&10^{-2}\{ h_A^2-0.6 
(h_V^0+{4\over 3}h_V^2) 
-0.0012 h_{n\Sigma^- K} 
-0.033(-h_V^{n\Sigma^- K^+} \\ \nonumber 
&&+{h_V^{p\Sigma^0 K^+}\over \sqrt{2}}) 
+0.2 (h^{pK}_A-h^{nK}_A) 
-0.006 h_{p\Lambda K} +0.088 h_V^{p\Lambda K^+} \\ 
&&-0.26 (|h^0_{\rho }|+{|h^2_{\rho }|\over \sqrt{6}}) 
-0.087 |h^1_{\omega }|+0.05 |h^1_{\phi }| \} 
 \; , 
\end{eqnarray} 
where we have set the phase of 
the vector meson contributions as discussed above, and used the relations in 
Eq. (\ref{eq:su2lim}). 
 
The expressions in Eqs. (\ref{RAS}-\ref{RAV}) illustrate the sensitivity of 
the radiative 
corrections to the various PV hadronic couplings. As expected on general 
grounds, the 
overall scale of $R_\sst{A}^{T=0,1}$ is at about the one percent level [see Eq. 
(\ref{eq:rascale})]. 
In terms of the conventional PV couplings, $\ratez$ is most sensitive to 
$h_\pi$ and 
$h_\rho^1$, while $\rateo$ is most strongly influenced by 
$h_\rho^0+h_\rho^2/\sqrt{6}$. 
The corrections also display strong dependences on the couplings 
$h_{V,A}^i$ not included 
in the standard analysis of nuclear and hadronic PV. In particular, the 
couplings $h_A^2$ 
and $h_V^0+4 h_V^2/3$ are weighted heavily in $\rateo$. In general, the 
sensitivity to 
the PV $NYK$ couplings is considerably weaker than the sensitivity to the 
$NN\pi$ and 
$NN\rho$ couplings. 
 
In order to make an estimate of $R_\sst{A}^{T=0,1}$, we require inputs 
for the PV 
couplings. To that end, we use the \lq\lq best values" for $h_\pi$, 
$h_{\rho}^i$, and 
$h_{\omega}^i$ given in Ref. \cite{fcdh}. These values are consistent with 
the fit of 
Ref. \cite{haxton}. For the $h_\phi^i$ we use the \lq\lq best values" of 
Ref. \cite{ddh}. 
The analyses given in Refs. \cite{ddh,haxton,fcdh}, together with 
experimental input, also allow 
for the standard couplings to take on a range of values. For example, the 
ranges for the 
$h_\omega^i$ given in Refs. \cite{ddh,fcdh} correspond to 
\begin{equation} 
-33 \le h_\omega^0+h_\omega^1 
\le 13 \ \ \ . 
\end{equation} 
In order to maintain consistency with the experimental constraint of 
Eq. (\ref{eq:ppcons}), one then requires 
\begin{equation} 
0\le h_\rho^0+h_\rho^1+h_\rho^2/\sqrt{6} 
\le -45 \ \ \ . 
\end{equation} 
We adopt this range even though it is smaller than the range given in 
Ref. \cite{fcdh}. Indeed, allowing the $h_\rho^i$ to assume the full ranges 
given in 
Ref. \cite{fcdh} would require the $h_\omega^i$ to vary outside their 
corresponding theoretical 
\lq\lq reasonable ranges" if the constraint of Eq. (\ref{eq:ppcons}) is to 
be satisfied. 
Since one expects $|h_\rho^1| << |h_\rho^{0,2}|$ \cite{ddh,fcdh}, 
we have a reasonable range of values for the important isoscalar $\rho$ 
contribution in 
Eq. (\ref{RAV}), and the rather broad range of values allowed for the 
$h_\rho^i$ contributes 
significantly to our estimated uncertainty in $\rateo$. For 
$h_{\omega,\phi}^i$, we use the ranges of 
Refs. \cite{ddh,fcdh}\footnote{Allowing the $h_\omega^i$ to assume positive 
values would 
require a sign change on the correspnding terms in Eqs. 
(\ref{RAS},\ref{RAV}).}. In contrast to 
the situation with the $h_\rho^0$ contribution, however, the variation in the 
$h_{\omega,\phi}^i$ over 
their \lq\lq reasonable ranges" has negligible impact on our estimated 
theoretical uncertainty. 
 
Estimating values for the Yukawa couplings $h_{NYK}$ and for the $h_{V,A}$ 
is more 
problematic---to date, no calculation on the level of Ref. \cite{ddh} has 
been performed 
for such couplings. Estimates for $h_{V,A}$, based on 
dimensional and factorization arguments,  were given in Ref. 
\cite{kaplan} and generally yielded 
values for 
$h_{V,A}$ in the non-strange sector on the order of $g_\pi$. 
For our 
central values, then, we take $h_{V,A}^i=g_\pi$, resulting in roughly 1\% 
contributions 
from the PV vector and axial vector interactions. Without performing a 
detailed 
calculation as in Ref. \cite{ddh}, one might also attempt to determine 
reasonable ranges for these 
parameters by looking to phenomenology. To that end, the authors of Ref. 
\cite{kaplan} considered 
analogies between the axial vector PV operators of Eq. (\ref{n2}-\ref{n3}) and 
contact operators needed to explain the size of $\Delta I=1/2$ hyperon 
P-wave decay 
amplitudes. From this analogy, these authors conclude that $|h_A^i| \sim 10 
g_\pi$ may 
be reasonable. However, whether such large ranges are consistent with 
nuclear PV 
data remains to be 
determined. In the absence of such an analysis, which goes beyond the scope 
of the present work, 
we adopt the range $-10 g_\pi \leq h_A^i \leq 10 g_\pi$ suggested in Ref. 
\cite{kaplan}. The 
corresponding uncertainties in the $R_A^{T=0,1}$ are roughly $\pm 10\%$. 
 
The implications of phenomenology for the $h_V^i$ are even less clear than 
for the $h_A^i$. 
However, we note that large values $h_V^i\sim \pm 10 g_\pi$ do not appear 
to be ruled out 
by hadronic and nuclear PV data. At tree-level, for example, the vector 
terms in 
${\cal L}^{\pi N}_{\Delta T=0, 1,2}$ do not contribute to the PV NN 
interaction through the 
one $\pi$-exchange amplitudes of Figure 8a. It is straightforward to show 
that the corresponding 
amplitude vanishes for on-shell nucleons \footnote{The on-shell 
approximation is generally used 
in deriving the PV NN potential from Feynman diagrams.}. Thus, at this 
level, purely hadronic 
PV processes are insensitive to the $h_V^i$ and provide no constraints on 
these couplings. 
In PV electromagnetic processes, however, the $h_V^i$ do contribute through 
PV two-body currents, 
such as those shown in Figure 8b. Nevertheless, one expects the impact of PV 
two-body currents to 
be considerably weaker than that of the PV NN potential. The PV 
$\gamma$-decay of $^{18}$F, 
for example,  is dominated by the mixing of a nearly-degenerate pair of 
$(J^\pi, T) = (0^-,0)$ and $(0^+, 1)$ states. The small energy denominator 
associated with 
this parity-mixing enhances the relative importance of the PV NN potential 
by roughly two orders 
of magnitude over the generic situation with typical nuclear level 
spacings. By contrast, the 
PV two-body currents do not participate in parity-mixing and receive no 
such enhancements. A similar 
situation holds for PV electromagnetic processes in other nuclei of interest. 
Hence, we expect the PV $\gamma$-decays of light nuclei to be relatively 
insensitive to the $h_V^i$, 
even if the latter are on the order of $10 g_\pi$. Consequently, we rather 
generously take 
$-10 g_\pi \leq h_V^0+4h_V^2/3 \leq 10 g_\pi$, yielding a $\pm 7\%$ 
contribution to the 
uncertainty in $\rateo$.  Allowing similarly large ranges for the PV $NYK$ 
couplings has a negligible 
impact on the uncertainty in the $R_A^{T=0,1}$. 
 
With these input values for the PV couplings, we arrive at the anapole 
contributions to 
$R_\sst{A}^{T=0,1}$ shown in Table I. The latter must be added to the 
one-quark Standard Model 
contributions, also shown in Table I. We compute the one-quark corrections 
using the on-shell 
parameters given in Refs. \cite{pdg,pdg96}. We emphasize that the quoted values 
for the $R_A^{T=0,1}$ 
are renormalization scheme-dependent. The relative size of the isovector 
one-quark 
corrections are smaller, for example, in the 
$\overline{\hbox{MS}}$ scheme, where one has $\rateo({\hbox{SM}})=-0.18$ and 
$\ratez({\hbox{SM}})=0.07$. The corresponding tree-level amplitude, 
however, is also smaller by a factor of $\sim 1.44$ 
than the on-shell tree-level amplitude. A reader working in the 
$\overline{\hbox{MS}}$ scheme should, therefore, take care to adjust the 
tree-level 
amplitude and SM radiative corrections appropriately from the on-shell 
values used here. 
Moreover, the anapole contributions to the $R_\sst{A}^{(T)}$ will be a 
factor of 
1.44 larger in the $\overline{\hbox{MS}}$ scheme since the tree-level 
amplitude is 
correspondingly smaller\footnote{Note that $\ratez$ gives the ratio of the 
isoscalar, 
axial vector amplitude to the tree-level isovector, axial vector amplitude. 
The sign 
of $\ratez$ as defined here is opposite that of Ref. \cite{MRM94}.}. 
 
Adding the one-quark and anapole contributions yields 
a large, negative value for $\rateo$. This result contains considerable 
theoretical uncertainty, 
mostly due to our liberal assignment of reasonable ranges to the $h_{V,A}$. 
Even with this generous 
theoretical uncertainty, however, $\rateo$ is still roughly a factor of two 
away from the apparent 
SAMPLE result. Compared with the one-quark SM contribution, the many-quark 
anapole contribution 
is relatively small -- though it does push the total in the right 
direction. The isoscalar 
correction, $\ratez$, is considerably smaller in magnitude than $\rateo$ 
yet retains a sizeable 
theoretical uncertainty. 
 
\begin{table}\label{tab1} 
\begin{center}~ 
\begin{tabular}{|c||c|c|}\hline 
Source & $\rateo$ & $\ratez$ 
\\ \hline\hline 
One-quark (SM)    & $-0.35$  & $0.05$ \\ \hline 
 Anapole  & $-0.06\pm 0.24$ & $0.01 \pm 0.14$ \\ \hline 
Total  &  $-0.41\pm 0.24$   &   $0.06\pm 0.14$ \\ \hline 
\end{tabular} 
\end{center} 
\caption{ One-quark Standard Model (SM) and many-quark anapole 
contributions to $V(A)\times A(N)$ radiative corrections. 
Values are computed in the 
on-shell scheme using $\sstw=0.2230$ .} 
\end{table} 
\section{Conclusions} 
\label{sec6} 
 
 
In view of the preliminary SAMPLE result for PV quasielastic electron 
scattering from $^2$H, we have 
up-dated our previous calculation of the axial vector radiative corrections 
$R_A^{T=0,1}$. 
Using the framework of HBChPT, we have computed all many-quark anapole 
contributions through 
${\cal O}(1/\lamchis)$. We include new one-loop contributions involving the 
PV vector couplings, 
$h_V^i$ and estimate the scale of the analytic, low-energy constants using 
resonance saturation. 
We fix the sign of the latter using the phenomenology of PV ${\vec p} p$ 
scattering and of 
nucleon EM form factors. We also show that large classes of loops involving 
decuplet intermediate states, magnetic insertions, and PV EM insertions 
vanish through ${\cal O}( 
1/\lamchis)$. Finally, we extend the previous analyses to include SU(3) 
symmetry, and determine that 
the impact of kaon loops is generally negligible. 
In the end, we find that $\rateo$---though large and negative---is still 
a factor 
of two or so away from the suggestion that $\rateo\sim -1$ from the SAMPLE 
experiment. Even allowing for 
considerable theoretical uncertainty -- dominated by the PV couplings 
$h_{V,A}^i$ -- there remains 
a sizable gap between our result and the preliminary experimental value. 
 
There exist a number of possible additional contributions to 
$R_\sst{A}^{T=0,1}$ not considered 
here which may ultimately account for the apparent experimental result. The 
most obvious include 
higher-order chiral corrections. This appears, however, 
to be an unlikely source of 
large contributions. 
On general grounds, we expect the size of the ${\cal O}(1/\Lambda_\chi^3)$ 
contributions to be 
suppressed by $m/\lamchi$ relative to those considered here, where $m$ 
denotes a pseudoscalar mass. 
For kaon loops, this suppression factor 
is only $\sim 1/2$; however, at ${\cal O}(1/\lamchis)$ kaon loops generate 
at most a few percent 
contribution to $R_\sst{A}^{T=0,1}$. The suppression factor for the next 
order pionic contributions 
is closer to 0.1. Hence, it would be surprising if the next order in the 
chiral expansion could 
close the factor of two gap with experiment. 
 
More promising sources of sizeable contributions include $Z-\gamma$ box 
graph contributions, where 
the full tower of hadronic intermediate states is included, as well as 
parity-mixing in the deuteron 
wavefunction. At a more speculative level, one might also consider 
contributions from physics beyond 
the Standard Model. For example, the presence of an additional, relatively 
light neutral gauge boson might modify the SM $V(e)\times A(q)$ amplitudes 
and contribute to $\rateo$. A popular class of $Z'$ models are generated by 
E$_6$ symmetry \cite{Lon86}. 
The contribution of an extra, neutral weak 
E$_6$ gauge boson $Z'$ is given by 
\begin{equation} 
\label{eq:e6boson} 
R_\sst{A}^{T=1}({\hbox{new}}) = {4\over 5} {1\over 1-4\sstw} \sin^2\phi 
{G_\phi'\over G_\mu}\ \ \ , 
\end{equation} 
where $\phi$ is a mixing angle which governs the structure of an additional 
U(1) group in E$_6$ theories \cite{Lon86} and $G_\phi'$ is the Fermi constant 
associated with the new U(1) group \cite{MRM99}. Note that this contribution 
has the wrong sign to account for the large negative value of $\rateo$. 
 
Alternatively, one might consider new tree-level interactions generated by 
supersymmetric extensions of the SM. Such interactions arise when R-parity, or 
equivalently, $B-L$, is not conserved ($B$ and $L$ denote baryon and lepton 
number, respectively). The contribution from R-parity violating SUSY 
interactions 
is given by \cite{Bar89,MRM99,MRM00} 
\begin{equation} 
\label{eq:rpv} 
\rateo({\hbox{new}}) =  \left({1\over 1-4x}\right) 
\left[ \Delta_{11k}^{\prime}(\tilde d_R^k) - \Delta_{1j1}^{\prime}(\tilde 
q^j_L) 
-\Delta_{12k}(\tilde e_R^k)(1-4x+4\lambda_x)\right] \ \ \ , 
\end{equation} 
where $x=\sstw$, 
\begin{equation} 
\label{eq:lamx} 
\lambda_x = {x(1-x)\over 1-2x} \left({1\over 1-\Delta r}\right)\sim 0.3\ \ \ , 
\end{equation} 
$\Delta r$ is a radiative correction, and where 
\begin{equation} 
\label{eq:rpv2} 
\Delta_{ijk}({\tilde f}) = {1\over 4\sqrt{2}} {|\lambda_{ijk}|^2\over G_\mu 
M_{\tilde f}^2 }\ \ \ , 
\end{equation} 
with $\tilde f$ denoting the superpartner of fermion $f$ and $i,j,k$ 
labeling fermion 
generations. The terms having a prime are semileptonic whereas the un-primed 
terms are purely leptonic. In principle, the correction in Eq. (\ref{eq:rpv}) 
could generate a negative contribution to $\rateo$. However, the various other 
electroweak data constrain the terms appearing in this expression. For example, 
relations between $G_\mu$ and other SM parameters require 
\begin{equation} 
-0.0023 \leq \Delta_{12k}(\tilde e_R^k) \leq 0.0028 \ \ \ , 
\end{equation} 
at 90 \% C.L., 
so that the first term in Eq. (\ref{eq:rpv}) cannot provide the large negative 
contribution needed to explain the SAMPLE result. Similarly, assuming only the 
semileptonic R-parity violating interactions modify the weak charge of nuclei, 
the recent determination of the cesium weak charge by the Boulder 
group\cite{Ben99,apv} 
implies that 
\begin{equation} 
0.0026 \le 2.6 \Delta_{11k}^{\prime}(\tilde d_R^k) - 2.9 
\Delta_{1j1}^{\prime}(\tilde q^j_L) \le 0.015\ \ \ , 
\end{equation} 
at 95 \% C.L. (for $m_H=300$ GeV). Thus, it appears unlikely that the 
second term in 
Eq. (\ref{eq:rpv}) could enhance $\rateo$ by a factor of two. 
 
In short, two of the most popular new physics scenarios having implications 
for low-energy phenomenology appear unlikely to enhance $\rateo$ significantly. 
Thus, if more conventional hadronic and nuclear processes cannot account for 
the SAMPLE result, one may be forced to consider more exotic alternatives. 
 
\section*{Acknowledgement} 
We wish to thank D. Beck, E.J. Beise, and R. McKeown for useful discussions. 
This work was supported in part under U.S. Department of Energy contract 
\#DE-AC05-84ER40150, the National Science Foundation 
and a National Science Foundation Young Investigator Award. 
 

\newpage 
\appendix 
 
\section{Formalism} 
In this section we first review the general parity and CP conserving 
Lagrangians including $N, \pi, \Delta, \gamma$ in the relativistic form. We 
follow 
standard conventions and introduce 
\begin{equation} 
\Sigma =\xi^2\  ,\ \  \xi =e^{i\pi\over F_\pi}\  ,\ \  \pi ={1\over 2} 
\pi^a \tau^a 
\end{equation} 
with $F_\pi =92.4$ MeV being the pion decay constant. 
The chiral vector and axial vector currents are given by 
\begin{eqnarray}\nonumber 
A_\mu &=& -{i\over 2}(\xi D_\mu \xi^\dag -\xi^\dag D_\mu 
\xi)=-{D_\mu\pi\over F_\pi} 
+O(\pi^3) \\ 
V_\mu &=&{1\over 2}(\xi D_\mu \xi^\dag +\xi^\dag D_\mu \xi)\ \ \ . 
\end{eqnarray} 
and we require also the gauge and chiral covariant derivativs 
\begin{eqnarray}\nonumber 
D_\mu\pi &=& \partial_\mu \pi -ie{\cal A}_\mu [Q,\pi]\\ 
{\cal D}_\mu &=& D_\mu +V_\mu\ \ \ , 
\end{eqnarray} 
with 
\begin{equation} 
Q=\left( \begin{array}{ll}{2\over 3}&0\\0&-{1\over 3} \end{array} \right) 
\end{equation} 
and ${\cal A}_\mu$ being the photon field. 
The chiral field strength tensors are 
\begin{equation} 
F^{\pm}_{\mu\nu}={1\over 2}(\partial_\mu{\cal A}_\nu-\partial_\nu{\cal A}_\mu) 
(\xi Q'\xi^\dag \pm \xi^\dag Q'\xi ) 
\end{equation} 
 with 
\begin{equation} 
Q'=\left( \begin{array}{ll}1&0\\0&0 
\end{array} \right) 
\end{equation} 
acting in the space of baryon isodoublets. 
 
For the moment, we restrict our attention to SU(2) flavor space and 
consider just $\pi$, $N$, and 
$\Delta$ degrees of freedom. We represent 
the nucleon as a two component isodoublet field, while for 
the $\Delta$, we 
use the isospurion formalism, treating the $\Delta$ field $T_\mu^i(x) $ as 
a vector spinor in both 
spin and  isospin space \cite{hhk} with the constraint $\tau^i T_\mu^i 
(x)=0$. The components of 
this field are 
\begin{equation} 
T^3_\mu =-\sqrt{{2\over 3}}\left( \begin{array}{l} \Delta^+\\ \Delta^0 
 \end{array} \right)_\mu\ \ \ , \ T^+_\mu =\left( \begin{array}{l} 
\Delta^{++}\\ 
\Delta^+/\sqrt{3} 
 \end{array} \right)_\mu \ \ \ , \ T^-_\mu =-\left( \begin{array}{l} 
\Delta^0/\sqrt{3}\\ 
\Delta^- 
 \end{array} \right)_\mu\ \ \ . 
\end{equation} 
The field $T^i_\mu$ also satisfies the constraints for the 
ordinary Schwinger-Rarita spin-${3\over 2}$ field, 
\begin{equation} 
\gamma^\mu T_\mu^i=0\ \ \  {\hbox{and}}\ \ \  p^\mu T_\mu^i=0\ \ \ . 
\end{equation} 
We eventually convert to the heavy baryon expansion, in which case the 
latter constraint 
becomes $v^\mu T_\mu^i=0$ with $v_\mu$ the heavy baryon velocity. 
 
It is useful to review the spacetime and chiral transformation properties 
of the various 
fields. Under a chiral transformation, 
\begin{eqnarray}\nonumber 
&\xi\to L\xi U^\dag=U\xi R^\dag\\ \nonumber 
& A_\mu \to U A_\mu U^\dag \\ 
&{\cal D}_\mu \to  U{\cal D}_\mu U^\dag  , 
\end{eqnarray} 
and 
\begin{equation} 
N\to UN\  ,\ \  T_\mu\to U T_\mu \  ,\ \  \Sigma\to L\Sigma R^\dag\ \ 
{\hbox{etc}}. 
\end{equation} 
 
In the $SU(2)$ sector parity violating effects are conveniently described 
by introducing the 
operators \cite{kaplan}: 
\begin{equation} 
X_L^a=\xi^\dag \tau^a \xi \  ,\ \    X_R^a=\xi \tau^a \xi^\dag \  ,\ \ 
X_{\pm}^a = X_L^a {\pm} X_R^a\ \ \ . 
\end{equation} 
which transform as 
\begin{equation} 
X_{L,R}^a\to U {\tilde {X_{L,R}^a}} U^\dag\ \ \ , 
\end{equation} 
with the index a rotating like a vector of $SU(2)_L$ and $SU(2)_R$ 
respectively. 
 
The P and CP transformation properties of these fields are shown in Table 2. 
 
\begin{table} 
\begin{center}~ 
\begin{tabular}{|c||c|c|}\hline 
Field & P & CP 
\\ 
\hline\hline 
$A_\mu$ & $-A^\mu$  & $-A^T_\mu$  \\ 
$N$ & $\gamma_0 N$ & $\gamma_0 C \bar N^T$ \\ 
$T_\mu$ & $- \gamma_0 T^\mu$ & $ -\delta (a)\gamma_0 C {\bar T}^{Ta}_\mu$ \\ 
$X_L^a$ & $X_R^a$ & $ \delta (a) X_{L}^{Ta}$ \\ 
$X_R^a$ & $ X_L^a$ & $ \delta (a) X_{R}^{Ta}$ \\ 
$F^{\pm}_{\mu\nu}$  & $\pm{F^{\pm}}^{\mu\nu}$ & $ -F^{\pm T\mu\nu}$ \\ 
&& \\ 
\hline 
\end{tabular} 
\end{center} 
\caption{\label{tab2} Parity (P) and CP transformation properties of 
chiral fields. Here, T denotes the transpose, $C$ is the charge 
conjugation matrix 
($C=i\gamma_2\gamma_0$ in the 
Dirac representation) and $\delta (i)=1, i=1, 3$  and $\delta (2)=-1$. 
} 
\end{table}

Finally, we note that in the Lagrangians of Section III, one has the following 
definitions: 
\begin{eqnarray}\nonumber 
{\cal D}_\mu^{ij} &=& \delta^{ij} {\cal D}_\mu -2i \epsilon^{ijk} V^k_\mu 
\\ \nonumber 
\omega_\mu^i &= & Tr[\tau^i A_\mu] \\ 
A_\mu^{ij} &= & \xi^{ik}_{3/2} A_\mu \xi^{kj}_{3/2}\ \ \ , 
\end{eqnarray} 
where 
$\xi^{ij}_{3/2} = {2\over 3}\delta^{ij}-{i\over 3} \epsilon_{ijk}\tau^k$ 
is the isospin $3/ 2$  projection operator. 
 
\section{The $SU(3)$ parity violating and CP conserving Lagrangian} 
\label{sec7} 
 
In this Appendix we list the parity violating and CP conserving $SU(3)$ 
Lagrangian for the pseudosclar meson octet and baryon octet. 
We are interested in the diagonal case of the parity violating electron 
nucleon scattering. Hence, 
we include only those interaction terms that ensure strangeness and charge 
conservation at each vertex. In the following 
we use $\xi =e^{i\pi\over F_\pi}, \pi ={1\over 2} \pi^a \lambda^a$, 
$X_L^a=\xi^\dag \lambda^a \xi$, $X_R^a=\xi \lambda^a \xi^\dag$, 
$X_{\pm}^a = X_L^a {\pm} X_R^a$, $[A, B]_{\pm}=AB \pm BA$. 
 
We classify the parity violating Lagrangian 
according to isospin violation $\Delta T =0, 1, 2$, which arises from the 
operators of $X_L^a, X_R^a$, their combinations and products. The $\Delta T=2$ 
piece comes from the operators ${\cal I}^{ab} \{X_L^a {\cal O} X_L^b \pm 
(L\leftrightarrow R)\}$ with ${\cal O}= N, \bar N, A_\mu$ and 
\begin{equation} 
{\cal I}^{ab}= 
{1\over 3}\left( \begin{array}{lll}1&0&0\\0&1&0\\0&0&-2 
\end{array} \right) \ \ \ , 
\end{equation} 
where $a, b=1, 2, 3$. Several 
operators contribute to the $\Delta T=1$ part, like $X_{\pm}^3, 
f^{3ab} \{X_L^a {\cal O} X_L^b \pm (L\leftrightarrow R)\}, 
d^{3ab} \{X_L^a {\cal O} X_L^b \pm (L\leftrightarrow R)\}$ where 
$f^{abc}, d^{abc}$ are the antisymmetric and symmetric structure 
constants of $SU(3)$ algebra. With the requirement that the final Lagrangian 
be hermitian, parity-violating and CP-conserving, the operator with $f^{3ab}$ 
vanishs. For the $\Delta T=0$ part relevant operators are ${\bf 1}, X_{\pm}^8, 
f^{8ab} \{X_L^a {\cal O} X_L^b \pm (L\leftrightarrow R)\}, 
d^{8ab} \{X_L^a {\cal O} X_L^b \pm (L\leftrightarrow R)\}, 
\delta^{ab}\{X_L^a {\cal O} X_L^b \pm (L\leftrightarrow R)\}$. For the same 
reason the $f^{8ab}$ structure does not contribute. Note the 
matrix identity $\lambda^a \lambda^b \lambda^a =4 (C_2({\bf 3}) -{1\over 2} 
C_2 ({\bf 8}) )\lambda^b$, where $C_2({\bf 3}), C_2 ({\bf 8})$ are the 
Casimir invariants of the basic and adjoint representations of $SU(3)$ group 
respectively. Hence, the operator containing $\delta^{ab}$ is identical to 
the unit operator. 
 
Based on these considerations, we obtain 
\begin{eqnarray}\label{l0}\nonumber 
{\cal L}^{\mbox{PV}}_{\Delta T =0} = 
h_3 F_\pi Tr { \bar N  [X_-^8, N]_+} 
+h_4 F_\pi Tr { \bar N  [X_-^8, N]_-} 
+v_1 Tr { \bar N \gamma^\mu [A_\mu, N]_+}&\\ \nonumber 
+v_2 Tr { \bar N \gamma^\mu [A_\mu, N]_-} 
+{v_7\over 2} Tr {\bar N \gamma^\mu A_\mu N X_+^8} 
+{v_8\over 2} Tr {\bar N \gamma^\mu  X_+^8 N A_\mu } &\\ \nonumber 
+{v_9\over 2} Tr {\bar N \gamma^\mu  [X_+^8, A_\mu]_+ N} 
+{v_{10}\over 2} Tr {\bar N \gamma^\mu N [X_+^8, A_\mu]_+ } 
+a_5 Tr {\bar N \gamma^\mu \gamma_5 A_\mu N X_-^8} &\\ \nonumber 
+a_6 Tr {\bar N \gamma^\mu \gamma_5 X_-^8 N A_\mu } 
+a_7 Tr {\bar N \gamma^\mu \gamma_5 [X_-^8, A_\mu]_+ N} 
+a_8 Tr {\bar N \gamma^\mu \gamma_5 N [X_-^8, A_\mu]_+ } &\\ \nonumber 
+\sqrt{3}v_{11} d^{8ab} Tr \{ \bar N\gamma^\mu N X_L^a A_\mu X_L^b + 
(L\leftrightarrow R) \} &\\ \nonumber 
+\sqrt{3}v_{12} d^{8ab} Tr \{ \bar N\gamma^\mu  X_L^a A_\mu X_L^b N + 
(L\leftrightarrow R) \} &\\ \nonumber 
+\sqrt{3}v_{13} d^{8ab} Tr \{ \bar N\gamma^\mu  X_L^a N [X_L^b, A_\mu]_+  + 
(L\leftrightarrow R) \} &\\ \nonumber 
+\sqrt{3}v_{14} d^{8ab} Tr \{ \bar N\gamma^\mu  [X_L^a, A_\mu]_+ N X_L^b  + 
(L\leftrightarrow R) \} &\\ \nonumber 
+\sqrt{3} a_9 d^{8ab} Tr \{ \bar N\gamma^\mu \gamma_5 N X_L^a A_\mu X_L^b - 
(L\leftrightarrow R) \} &\\ \nonumber 
+\sqrt{3}a_{10} d^{8ab} Tr \{ \bar N\gamma^\mu \gamma_5 X_L^a A_\mu X_L^b N - 
(L\leftrightarrow R) \} &\\ \nonumber 
+\sqrt{3}a_{11} d^{8ab} Tr \{ \bar N\gamma^\mu \gamma_5 X_L^a N [X_L^b, 
A_\mu]_+  - 
(L\leftrightarrow R) \} &\\ 
+\sqrt{3}a_{12} d^{8ab} Tr \{ \bar N\gamma^\mu \gamma_5 [X_L^a, A_\mu]_+ N 
X_L^b  - 
(L\leftrightarrow R) \} &\; , 
\end{eqnarray} 
 
\begin{eqnarray}\label{l1}\nonumber 
{\cal L}^{\mbox{PV}}_{\Delta T =1} = 
h_1 F_\pi Tr { \bar N  [X_-^3, N]_+} 
+h_2  F_\pi Tr { \bar N  [X_-^3, N]_-} 
+{v_3\over 2} Tr {\bar N \gamma^\mu A_\mu N X_+^3}&\\ \nonumber 
+{v_4\over 2} Tr {\bar N \gamma^\mu  X_+^3 N A_\mu } 
+{v_5\over 2} Tr {\bar N \gamma^\mu  [X_+^3, A_\mu]_+ N} 
+{v_6\over 2} Tr {\bar N \gamma^\mu N [X_+^3, A_\mu]_+ } &\\ \nonumber 
+a_1 Tr {\bar N \gamma^\mu \gamma_5 A_\mu N X_-^3} 
+a_2 Tr {\bar N \gamma^\mu \gamma_5 X_-^3 N A_\mu } 
+a_3 Tr {\bar N \gamma^\mu \gamma_5 [X_-^3, A_\mu]_+ N}&\\ \nonumber 
+a_4 Tr {\bar N \gamma^\mu \gamma_5 N [X_-^3, A_\mu]_+ } 
+v_{15} d^{3ab} Tr \{ \bar N\gamma^\mu N X_L^a A_\mu X_L^b + 
(L\leftrightarrow R) \} &\\ \nonumber 
+v_{16} d^{3ab} Tr \{ \bar N\gamma^\mu  X_L^a A_\mu X_L^b N + 
(L\leftrightarrow R) \} &\\ \nonumber 
+v_{17} d^{3ab} Tr \{ \bar N\gamma^\mu  X_L^a N [X_L^b, A_\mu]_+  + 
(L\leftrightarrow R) \} &\\ \nonumber 
+v_{18} d^{3ab} Tr \{ \bar N\gamma^\mu  [X_L^a, A_\mu]_+ N X_L^b  + 
(L\leftrightarrow R) \} &\\ \nonumber 
+ a_{13} d^{3ab} Tr \{ \bar N\gamma^\mu \gamma_5 N X_L^a A_\mu X_L^b - 
(L\leftrightarrow R) \} &\\ \nonumber 
+a_{14} d^{3ab} Tr \{ \bar N\gamma^\mu \gamma_5 X_L^a A_\mu X_L^b N - 
(L\leftrightarrow R) \} &\\ \nonumber 
+a_{15} d^{3ab} Tr \{ \bar N\gamma^\mu \gamma_5 X_L^a N [X_L^b, A_\mu]_+  - 
(L\leftrightarrow R) \} &\\ 
+a_{16} d^{3ab} Tr \{ \bar N\gamma^\mu \gamma_5 [X_L^a, A_\mu]_+ N X_L^b  - 
(L\leftrightarrow R) \} &\; , 
\end{eqnarray} 
 
\begin{eqnarray}\label{l2}\nonumber 
{\cal L}^{\mbox{PV}}_{\Delta T =2} = 
{v_{19}\over 2} {\cal I}^{ab} Tr \{ \bar N\gamma^\mu N X_L^a A_\mu X_L^b + 
(L\leftrightarrow R) \} &\\ \nonumber 
+ {v_{20}\over 2} {\cal I}^{ab}Tr \{ \bar N\gamma^\mu  X_L^a A_\mu X_L^b N + 
(L\leftrightarrow R) \} &\\ \nonumber 
+ {v_{21}\over 2} {\cal I}^{ab}Tr \{ \bar N\gamma^\mu  X_L^a N [X_L^b, 
A_\mu]_+  + 
(L\leftrightarrow R) \} &\\ \nonumber 
+ {v_{22}\over 2} {\cal I}^{ab}Tr \{ \bar N\gamma^\mu  [X_L^a, A_\mu]_+ N 
X_L^b  + 
(L\leftrightarrow R) \} &\\ \nonumber 
+ {a_{17}\over 2} {\cal I}^{ab}Tr \{ \bar N\gamma^\mu \gamma_5 N X_L^a 
A_\mu X_L^b - 
(L\leftrightarrow R) \} &\\ \nonumber 
+ {a_{18}\over 2} {\cal I}^{ab}Tr \{ \bar N\gamma^\mu \gamma_5 X_L^a A_\mu 
X_L^b N - 
(L\leftrightarrow R) \} &\\ \nonumber 
+ {a_{19}\over 2} {\cal I}^{ab}Tr \{ \bar N\gamma^\mu \gamma_5 X_L^a N 
[X_L^b, A_\mu]_+  - 
(L\leftrightarrow R) \} &\\ 
+ {a_{20}\over 2} {\cal I}^{ab}Tr \{ \bar N\gamma^\mu \gamma_5 [X_L^a, 
A_\mu]_+ N X_L^b  - 
(L\leftrightarrow R) \} &\; . 
\end{eqnarray} 
 
These Lagrangians contain 4 Yukawa couplings, 20 axial vector couplings and 
22 vector 
couplings, all of which should be fixed from the 
experimental data or from model 
calculations.  In reality, however, we have only limited information which 
constrains a few of them. It is useful to expand 
the above Lagrangians to the order involving the 
minimum number of Goldstone bosons 
and to collect those 
vertices needed in the calculation of $R_A$: 
 
\begin{eqnarray}\label{yuk}\nonumber 
{\cal L}^{1\pi}_{Yukawa} =2\sqrt{2}i (h_1+h_2) (\bar p n \pi^+ -\bar n p 
\pi^-)&\\ \nonumber 
+i[h_1-h_2+\sqrt{3}(h_3-h_4)](\bar p \Sigma^0 K^+-{\bar{\Sigma^0}} p K^-) 
&\\ \nonumber 
+\sqrt{2} i[h_1-h_2+\sqrt{3}(h_3-h_4)](\bar n \Sigma^- K^+-{\bar{\Sigma}^-} 
n K^-)&\\ 
-i[{h_1\over\sqrt{3}}+\sqrt{3}h_2+h_3+3h_4](\bar p \Lambda 
K^+-{\bar\Lambda} p K^-) 
+\cdots &\; . 
\end{eqnarray} 
 
\begin{eqnarray}\label{vec}\nonumber 
{\cal L}^{1\pi}_{V} =-{h_V^{pn\pi^+}\over F_\pi} \bar p\gamma^\mu n D_\mu\pi^+ 
-{h_V^{p\Sigma^0 K^+}\over F_\pi} \bar p \gamma^\mu\Sigma^0 D_\mu K^+ &\\ 
-{h_V^{n\Sigma^- K^+}\over F_\pi} \bar n \gamma^\mu\Sigma^- D_\mu K^+ 
-{h_V^{p\Lambda K^+}\over F_\pi} \bar p \gamma^\mu\Lambda D_\mu K^+ + h.c. 
+\cdots 
&\; , 
\end{eqnarray} 
where 
\begin{eqnarray}\nonumber 
&h_V^{pn\pi^+}={v_1+v_2\over \sqrt{2}}+{4\sqrt{2}\over 3}(v_{14}-v_{12}) 
+{\sqrt{6}\over 3}(v_7+v_9)+{\sqrt{2}\over 3} v_{20} \\ \nonumber 
&h_V^{p\Sigma^0 K^+}= 
{1\over 2}(v_1-v_2+v_4+v_6)+{v_8-v_{10}\over 2\sqrt{3}} 
+{2\over 3}(v_{11}-v_{13}-v_{15}-v_{21}-{1\over 2}v_{17})+2v_{18} \\ \nonumber 
&h_V^{n\Sigma^- K^+}={1\over \sqrt{2}}(v_1-v_2+v_6-v_4)+{1\over \sqrt{6}} 
(v_8-v_{10})+{\sqrt{2}\over 3}(v_{17}+v_{21})+{2\sqrt{2}\over 3} 
(v_{11}-v_{13}-v_{15})&\\ \nonumber 
&h_V^{p\Lambda K^+}={1\over\sqrt{3} } 
(-{v_1\over 2}+{2\over 3}v_{11}-{4\over 3}v_{12}+{16\over 3}v_{13} 
-{2\over 3}v_{14}-{2\over 3}v_{15}+{4\over 3}v_{16}-{17\over 3}v_{17} \\ 
&+{4\over 3}v_{18}-{3\over 2}v_2+{v_4\over 2}-v_5+{v_6\over 2}) 
+{v_8-v_{10}\over 6}+{2v_7+v_9\over 3}&\; . 
\end{eqnarray} 
 
\begin{eqnarray}\label{axial}\nonumber 
{\cal L}^{2\pi}_{A} =-i{h_A^{p\pi}\over f^2_\pi}\bar p\gamma^\mu 
\gamma_5 p 
(\pi^+ D_\mu \pi^--\pi^- D_\mu \pi^+)&\\ \nonumber 
-i{h_A^{pK}\over f^2_\pi}\bar p\gamma^\mu \gamma_5 p 
(K^+ D_\mu K^--K^- D_\mu K^+) &\\ \nonumber 
-i{h_A^{n\pi}\over f^2_\pi}\bar n\gamma^\mu \gamma_5 n 
(\pi^+ D_\mu \pi^--\pi^- D_\mu \pi^+)&\\ 
-i{h_A^{nK}\over f^2_\pi}\bar n\gamma^\mu \gamma_5 n 
(K^+ D_\mu K^--K^- D_\mu K^+) 
+\cdots &\; , 
\end{eqnarray} 
where 
\begin{eqnarray}\nonumber 
&h_A^{p\pi}=2a_3-{4\over 3}a_{16}+{2\over 
3}a_{14}-a_{18}+2a_{13} \\ \nonumber 
&h_A^{pK}=a_3-a_4+\sqrt{3}(a_7-a_8)+a_9+a_{10}+a_{11}+a_{12} 
+{1\over 3}(a_{16}+a_{14}-a_{18}+a_{13}-a_{19})\\ \nonumber 
&h_A^{n\pi}=2a_3 
-{4\over 3}a_{16}+{2\over 3}a_{14}+a_{18}+2a_{13}\\ 
&h_A^{nK}=a_4+\sqrt{3}a_8+a_9+{5\over 2}a_{10}-2a_{11}-a_{15} 
+a_{14}+{1\over 3}(a_{18}+a_{19}+a_{13}) 
\end{eqnarray} 
Note only $a_{18}$ contributes 
to $R_A$ in the parity violating two pions vertices. In the two kaons vertices 
$a_{3-4}, a_{7-8}, a_{10-19}$ all lead to nonzero contribution to $R_A$. 
 
\section{$\Delta$ Intermediate States and EM insertions} 
 
 
As noted in Section IV, the amplitudes of Figures 4-6 vanish through ${\cal 
O}(1/{\Lambda^2_\chi})$. 
Below, we briefly summarize the reasons behind this result. 
 
 
\subsection{PV $\pi \Delta N$ contribution} 
 
 
In the case where the $\Delta$ enters as an intermediate state we have the 
Feynman 
diagrams shown in Figure 4. 
Since the final and initial states are both nucleons, the two-pion parity 
violating vertices in Eqs. (\ref{d1})-(\ref{d3}) arise first at two-loop order 
and contribute to the 
nucleon  anapole moment at the order of $O({1/ \Lambda_{\chi}^3})$. Although 
the PV $N\Delta\pi$ 
interactions nominally contribute at lower order, in this case such 
contributions vanish up to ${\cal 
O}(1/\lamchis)$. The reason is as follows. Each of  the parity violating 
and CP conserving single 
pion vertices has the same Lorentz structure---$i\gamma_5$. In the heavy 
baryon expansion, the relevant vertices are obtained by the substitution 
$P_+ i\gamma_5 P_+$, which vanishes. The leading nonzero contribution arises 
at first order in 
the $1/\mn$ expansion. Consequently, its contribution to the nucleon 
anapole moment appears only 
at ${\cal O}({1/\Lambda_{\chi}^2 m_N})$, and since in this work we truncate at 
${\cal 
O}(1/\lamchis)$,  the PV $\pi \Delta N$ vertices do not contribute. 
 
\subsection{Magnetic moment insertions} 
 
The nucleon has a large isovector magnetic moment. We 
thus consider associated 
possible PV 
chiral loop corrections which lead to a nucleon anapole moment. 
The relevant diagrams are shown in Figure 5. 
At $O({1/\Lambda_{\chi}^2})$ there are only four 
relevant diagrams Figures 5a-d. Since the 
magnetic moment is of $O({1/\Lambda_{\chi}})$ and the strong pion baryon 
vertex 
is of $O({1/ F_\pi})$, the remaining PV vertex must be a Yukawa coupling if 
the loop is to 
contribute at ${\cal O}(1/\lamchis)$ or lower. 
For the nucleon magnetic moment insertion we have, for example, 
\begin{equation}\label{pp} 
iM_{5a} + iM_{5b}=i\epsilon^{\mu\nu\alpha\beta}\varepsilon_\mu q_\nu v_\alpha 
{\sqrt{2}g_A e\mu_N h_{\pi N}\over m_N F_\pi} [S_\beta, S_\sigma]_+ 
\int {d^Dk\over (2\pi)^D} 
{k^\sigma\over v\cdot k} {1\over v\cdot (q+k)}{1\over k^2-m_\pi^2 
+i\epsilon} \;, 
\end{equation} 
where $e_\mu$ is the photon polarization vector and $\mu_N$ is the 
nucleon magnetic moment. The denominator of the integrand in (\ref{pp}) is 
nearly the same as 
for $M_{3e}$. The numerator contains a single factor of $S\cdot k$. Hence, 
Figures 5a and 5b vanish for the same reason as does $M_{3e}$. 
 
For the nucleon delta transition magnetic moment insertion we have 
\begin{eqnarray}\label{p1} \nonumber 
iM_{5c}+iM_{5d}&=&-{2\over \sqrt{3}}{ g_{\pi N\Delta} e\mu_{\Delta N} 
h_{\pi N}\over 
m_N F_\pi} 
(q_\sigma \epsilon_\nu -\epsilon_\sigma q_\nu) 
[P^{\mu\nu}_{3/2} S^\sigma + S^\sigma P^{\nu\mu}_{3/2}]\\ 
&&\ \ \ \ \times \int {d^Dk\over (2\pi)^D} 
{k_\mu \over v\cdot k} {1\over v\cdot (q+k)}{1\over k^2-m_\pi^2 +i\epsilon} \;, 
\end{eqnarray} 
where $\mu_{\Delta N}$ is the nucleon delta transition magnetic moment and 
$P^{\mu\nu}_{3/2}=g^{\mu\nu}-v^\mu v^\nu +{4\over 3} S^\mu S^\nu$ is the 
spin ${3\over 2}$ projection operator in the heavy baryon chiral perturbation 
framework. Since the integrand is the same as in $M_{3e}$ the integral is 
proportional to $v_\mu$. 
Moreover, $v_\mu P^{\mu\nu}_{3/2}=0$, so that $M_{5c}+M_{5d}=0$. Finally, the 
$\Delta$ magnetic insertions of Figures 5e-h 
require the PV $N\Delta\pi$ vertex, which starts off at ${\cal O}(1/\mn 
F_\pi)$. Thus, the latter 
do not contribute up to ${\cal O}(1/\lamchis)$. In short, none of the 
magnetic insertions contribute at the order to which we work in this analysis. 
 
\subsection{PV electromagnetic insertions} 
 
Another possible source to the nucleon anapole moment arises from the 
PV magnetic moment like insertions as shown in Figure 6. 
All three PV $\gamma NN$ vertices $c_{1-3}$ in Eq. (\ref{n4}) and PV 
$\gamma\Delta N$ vertices 
$d_{4-6}$ in Eq. (\ref{d4}) start off with one pion, so they are of order 
${\cal O}({1/ 
\Lambda_{\chi} F_\pi})$. 
Vertices $d_{7,8}$ have two pions and are order ${\cal O}({1/\Lambda_{\chi} 
F_\pi^2})$. 
The corresponding contributions to the nucleon anapole moment appear at 
order of 
${\cal O}({1/ \Lambda_{\chi}^3 })$ or ${\cal O}({1/ 
\Lambda_{\chi}^4})$, respectively. 
The leading PV $\gamma\Delta N$ vertices $d_{1-3}$ do not have pions and are 
of the order ${\cal O}({1/ \Lambda_{\chi} })$. In our case, however, 
the final and initial 
states are both nucleons. The $\Delta$ can appear as the intermediate state 
inside 
the chiral loop, which leads to an additional factor ${1/ F_\pi^2}$ 
from two 
strong vertices. In the end $d_{1-3}$ contributes to the nucleon anapole moment 
at  ${\cal O}({1/ \Lambda_{\chi}^3 })$. Finally, the PV $\gamma \Delta\Delta$ 
vertices contain one $\pi$. Since the $\Delta$ can only appear as an 
intermediate 
state, this vertex contributes at two-loop order and is of higher-order in 
chiral counting than we consider here (the corresponding diagrams are not 
shown). Thus, to ${\cal O}(1/\lamchis)$, the PV electromagnetic insertions 
do not contribute. 
 
\newpage 
 
{\bf Figure Captions} 
\begin{center} 
{\sf 
FIG 1.} {Axial vector $\gamma NN$ coupling, generated by PV hadronic 
interactions} 
\end{center} 
\begin{center} 
{\sf 
Figure 2.} {Feynman diagrams for polarized electron nucleon scattering. 
Figure 2a gives 
tree-level $Z^0$-exchange amplitude, while FIG. 2b gives the anapole moment 
contribution. 
The dark circle indicates an axial vector coupling.} 
\end{center} 
\begin{center} 
{\sf 
Figure 3.} {Meson-nucleon intermediate state contributions to the 
nucleon anapole moment.  The shaded circle denotes the PV vertex. 
The solid, dashed and curly lines correspond to the nucleon, pion and 
photon respectively. 
For the $SU(3)$ case the intermediate states can also be hyperons and kaons.} 
\end{center} 
\begin{center} 
{\sf 
Figure 4.} {The contribution to the nucleon anapole moment from PV $\pi 
\Delta N$ vertices. The double line 
is the 
$\Delta $ intermediate state.  } 
\end{center} 
\begin{center} 
{\sf 
Figure 5.} {Anapole moment contributions generated by insertions of the 
baryon magnetic moment operator, 
denoted by the cross, and the PV hadronic couplings, denoted by the shaded 
circle.} 
\end{center} 
\begin{center} 
{\sf 
Figure 6.} {PV electromagnetic 
insertions, 
denoted by the overlapping cross and shaded circle.} 
\end{center} 
\begin{center} 
{\sf 
Figure 7.} {Vector meson contribution to the anapole moment. Shaded circle 
indicates PV 
hadronic coupling.} 
\end{center} 
\begin{center} 
{\sf 
Figure 8.} {Contributions to (a) PV NN interaction and (b) PV two-body 
current generated by 
the vector terms in Eqs. (\ref{n1}-\ref{n3}).} 
\end{center}

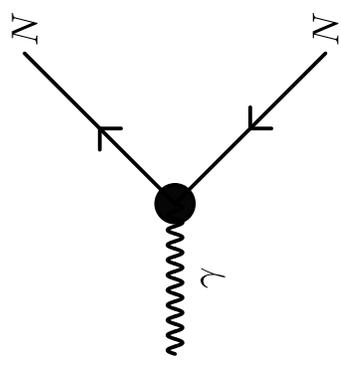
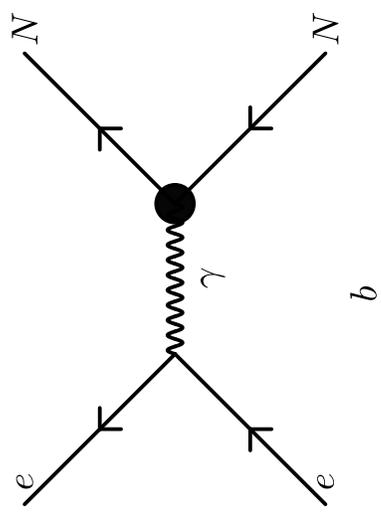
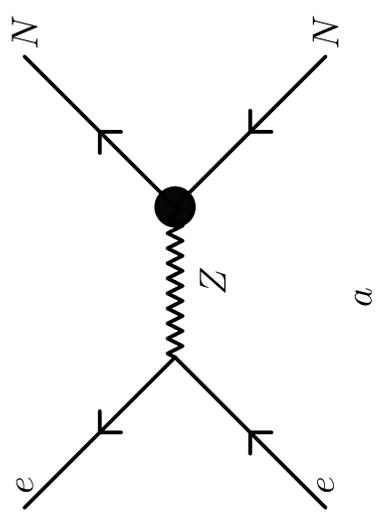
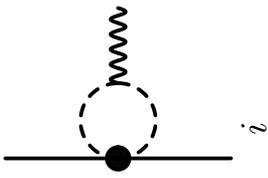
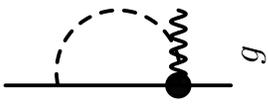
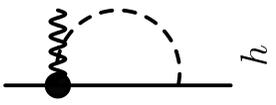
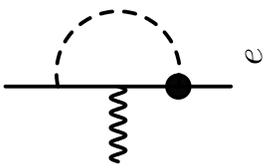
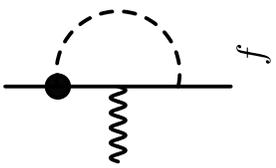
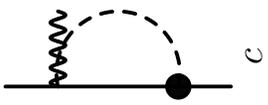
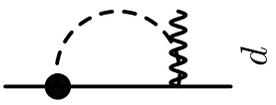
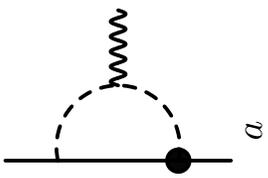
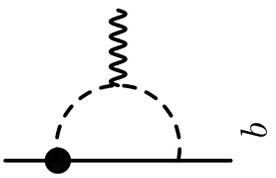
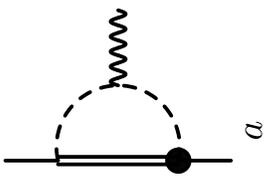
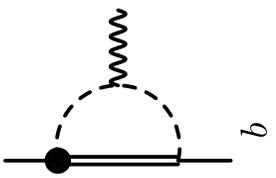
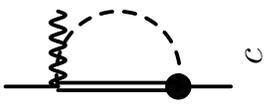
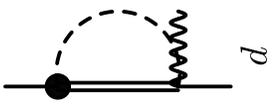
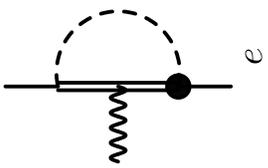
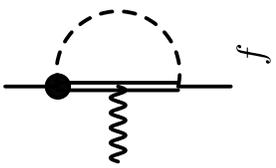
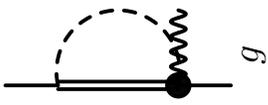
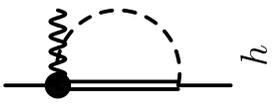
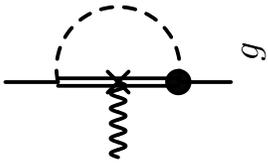
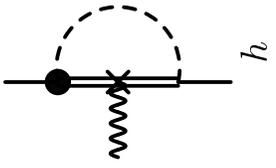
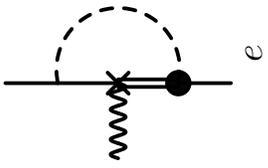
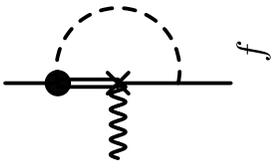
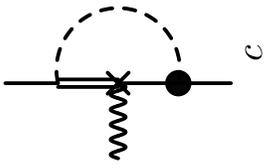
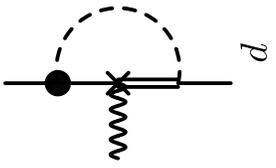
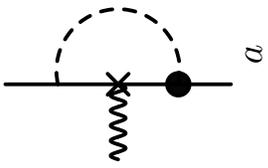
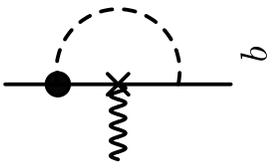
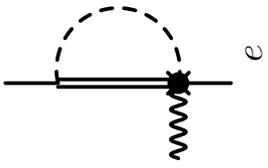
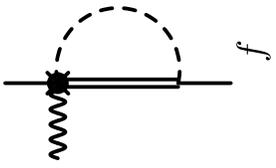
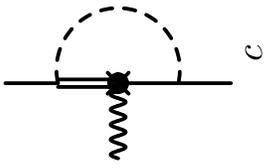
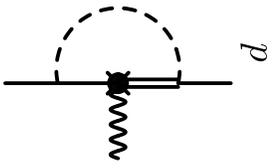
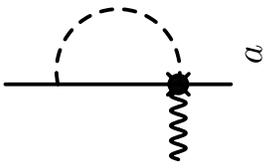
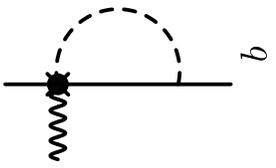
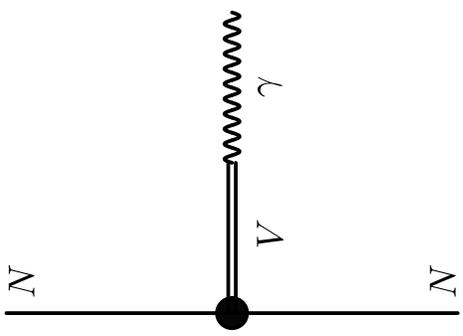
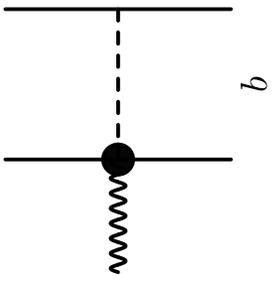
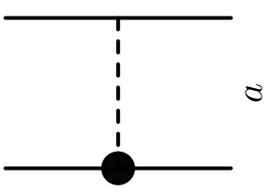
\end{document}